\documentclass[twocolumn,tighten]{aastex62}
\pdfoutput=1 
\usepackage{amsmath,amstext}
\usepackage[T1]{fontenc}
\usepackage{apjfonts} 
\usepackage[figure,figure*]{hypcap}
\usepackage{cleveref}
\usepackage{xcolor}
\usepackage{longtable}
\usepackage{enumitem}
\usepackage{float}

\newcommand*{\hmpc}{h^{-1}\textrm{Mpc}}
\newcommand*{\invhmpc}{h~ \textrm{Mpc}^{-1}}
\newcommand*{\hkpc}{h^{-1}\textrm{kpc}}
\newcommand*{\hgpc}{h^{-1}\textrm{Gpc}}
\newcommand{\hmsun}{h^{-1}\textrm{M}_{\odot}}

\newcommand{\nbody}{$N$-body}

\newcommand\aemulus{{\sc Aemulus}}

\shorttitle{Aemulus I: Simulations}
\shortauthors{DeRose et al.}

\begin{document}

\title{The Aemulus Project I: Numerical Simulations for Precision Cosmology}

\author{Joseph DeRose}
\affiliation{Kavli Institute for Particle Astrophysics and Cosmology and Department of Physics, Stanford University, Stanford, CA 94305, USA}
\affiliation{Department of Particle Physics and Astrophysics, SLAC National Accelerator Laboratory, Stanford, CA 94305, USA}

\author{Risa H. Wechsler}
\affiliation{Kavli Institute for Particle Astrophysics and Cosmology and Department of Physics, Stanford University, Stanford, CA 94305, USA}
\affiliation{Department of Particle Physics and Astrophysics, SLAC National Accelerator Laboratory, Stanford, CA 94305, USA}

\author{Jeremy L. Tinker}
\affiliation{Center for Cosmology and Particle Physics, Department of Physics, New York University, 4 Washington Place, New York, NY 10003, USA}
 
\author{Matthew R. Becker}
\affiliation{Kavli Institute for Particle Astrophysics and Cosmology and Department of Physics, Stanford University, Stanford, CA 94305, USA}
\affiliation{Department of Particle Physics and Astrophysics, SLAC National Accelerator Laboratory, Stanford, CA 94305, USA}
\affiliation{Civis Analytics, Chicago, IL 60607, USA}

\author{Yao-Yuan Mao}
\affiliation{Department of Physics and Astronomy and the Pittsburgh Particle Physics, Astrophysics and Cosmology Center (PITT PACC), University of Pittsburgh, Pittsburgh, PA 15260, USA}

\author{Thomas McClintock}
\affiliation{Department of Physics, University of Arizona, Tuscon, AZ 85721, USA}

\author{Sean McLaughlin}
\affiliation{Kavli Institute for Particle Astrophysics and Cosmology and Department of Physics, Stanford University, Stanford, CA 94305, USA}
\affiliation{Department of Particle Physics and Astrophysics, SLAC National Accelerator Laboratory, Stanford, CA 94305, USA}

\author{Eduardo Rozo}
\affiliation{Department of Physics, University of Arizona, Tuscon, AZ 85721, USA}

\author{Zhongxu Zhai}
\affiliation{Center for Cosmology and Particle Physics, Department of Physics, New York University, 4 Washington Place, New York, NY 10003, USA}

\begin{abstract}
The rapidly growing statistical precision of galaxy surveys has lead to a need for ever-more precise predictions of the observables used to constrain cosmological and galaxy formation models. The primary avenue through which such predictions will be obtained is suites of numerical simulations. These simulations must span the relevant model parameter spaces, be large enough to obtain the precision demanded by upcoming data, and be thoroughly validated in order to ensure accuracy. In this paper we present one such suite of simulations, forming the basis for the \aemulus\ Project, a collaboration devoted to precision emulation of galaxy survey observables. We have run a set of 75 $(1.05 ~\hgpc)^3$ simulations with mass resolution and force softening of $3.51\times 10^{10} \left(\frac{\Omega_{m}}{0.3}\right) ~\hmsun$ and $20~\hkpc$ respectively in 47 different $w$CDM cosmologies spanning the range of parameter space allowed by the combination of recent Cosmic Microwave Background, Baryon Acoustic Oscillation and Type Ia Supernovae results. We present convergence tests of several observables including spherical overdensity halo mass functions, galaxy projected correlation functions, galaxy clustering in redshift space, and matter and halo correlation functions and power spectra. We show that these statistics are converged to $1\%$ ($2\%$) for halos with more than 500 (200) particles respectively and scales of $r>200 ~ \hkpc$ in real space or $k\sim 3 ~ \invhmpc$ in harmonic space for $z\le 1$. We find that the dominant source of uncertainty comes from varying the particle loading of the simulations. This leads to large systematic errors for statistics using halos with fewer than 200 particles and scales smaller than $k\sim4~ \invhmpc$. We provide the halo catalogs and snapshots detailed in this work to the community at \url{https://AemulusProject.github.io}.
\end{abstract}


\section{Introduction}

The era of precision cosmology from galaxy surveys is upon us. Galaxy survey data sets have achieved comparable constraining power on a subset of cosmological parameters to measurements of the Cosmic Microwave Background (CMB) \citep{Alam2017,Y13x2}, but unlike the CMB, these constraints rely on the measurement and modeling of non-linear structure. In a very real sense these analyses are already systematics limited, disregarding significant portions of their data in order to mitigate modeling uncertainties. For example, \citet{Y13x2} limited itself to scales for which baryonic feedback and non-linear effects from galaxy biasing could be ignored. \citet{Alam2017}, presenting the final analysis of the BOSS galaxy redshift survey, restricted their redshift space distortion measurements to $s>20 ~\hmpc$ and $k<0.15 ~\invhmpc$ in configuration and Fourier space respectively to avoid uncertainties in modeling the galaxy velocity field.

Analytic models of these effects for simply selected samples are improving, but even the best models only claim to be accurate to the percent level at $k\sim 0.3~ \invhmpc$ for matter and halo power spectra before taking into account effects due to hydrodynamics, feedback and redshift space distortions \citep{Cataneo2017, Perko2016}. Non-linear effects are much more difficult to avoid in the halo mass function (HMF), and analytic predictions such as those in \citet{PressSchechter74} and \citet{ShethTormen99} are only accurate at the $\sim 10\%$ level \citep{Tinker2008}. Depending on the observable, this level of precision is either already a dominant source of error, or will be in the very near future \citep[see e.g.][]{Tinker2012}. While $1\%$ precision in observables is often quoted as a necessary goal, the required precision on predictions for observables is often not this stringent. For instance, \citet{McClintock2018} determines that the precision required for the halo mass function in order for it to contribute no more than $10\%$ of the total uncertainty in cluster mass calibration for upcoming surveys is $3\%$ at its most demanding.

While analytic methods struggle with non-linear structure formation, a clear alternative exists in numerical simulations. In the case of gravity, where we have a well-understood standard theory described by General Relativity, the effectiveness of simulations is limited only by the coarseness of the discretization allowed by currently available computers. Different algorithms for solving for non-linear structure growth in dark matter only simulations have been shown to produce predictions for the matter power spectrum that are converged at better than the $1\%$ level to $k\sim 1 ~\invhmpc$ \citep{Heitmann2009, Schneider2016}. It should be noted that these studies are of relative convergence, whereas studies of absolute convergence to the true physical solution is still an open question that likely depends on a better understanding of baryonic physics, neutrinos, and the nature of dark matter itself. Because of the relative successes of the aforementioned simulations, almost all cosmological analyses involving galaxy surveys now use them in some form \citep{y1sim2params,Kitaura2016,Joudaki2018}.

While great strides have been made in improving their computational efficiency, \nbody\ simulations are still relatively expensive. For example the $\textsc{ds14\_a}$ simulation \citep{Skillman2014}, one of the largest simulations run to date with a simulated volume of $(8~\hgpc)^3$ and $1.07\times10^{12}$ particles, took approximately 34 hours on 12,288 nodes, approximately $2/3$ of the \textsc{Titan} supercomputer. While this simulation approaches the volume of many ongoing and upcoming galaxy surveys, it does not resolve even all of the host halos of galaxies in a survey like DES. 

Cosmological parameter constraints typically rely on sampling schemes such as Monte Carlo Markov Chains (MCMC) in order to explore parameter space. Modern analysis including cosmological and nuisance parameters numbering in the tens must sample on the order of millions of different cosmologies in order to reach convergence. Running an \nbody\ simulation at each of these steps is not a prospect that will be achievable in the near future, even when considering smaller simulations than $\textsc{ds14\_a}$, such as those presented in this work. Thus, there is a need for methodologies which can use relatively few simulations to make robust predictions for the full cosmological parameter space being constrained. Much of the work in this area has been driven by the need for accurate predictions of the matter power spectrum for weak lensing analyses. For example, the \textsc{Halofit} methodology \citep{Smith02,Takahashi2012} fit an analytic expression to a set of \nbody\ simulations in various cosmologies to obtain predictions for the matter power spectrum accurate to $5\%$ for $k<1~\invhmpc$ and $10\%$ for $1~\invhmpc<k<10~\invhmpc$.

Investigations into more advanced methodologies are ongoing, typically combining algorithms for optimally sampling a chosen cosmological parameter space and a method for interpolating between the observables at the sampled cosmologies. This approach, dubbed cosmic emulation, was first demonstrated for the matter power spectrum in \cite{Heitmann2009}. They showed convergence of their simulation results with respect to a number of choices made in solving the \nbody\ problem, including mass resolution, force softening and simulation volume. This work has since been extended to the Friends-of-Friends halo mass function \citep{Heitmann2016}, galaxy correlation function and galaxy-shear cross correlation function \citep{Wibking2017}, among other observables. Studies of the convergence of these statistics are not as complete as those for the matter power spectrum. Work towards validating the convergence of these statistics is vital to ensuring the accuracy of predictions built from simulations. 

This type of validation is the primary concern of this work. The simulations presented here form the basis for the first set of emulators that is being built as a part of the \aemulus\ project, a collaboration focused on the emulation of galaxy survey observables. The goal of the validation presented here is to provide robust convergence estimates for the statistics in question so that they may be properly accounted for in emulators built from these simulations. Emulators for the halo mass function and redshift space galaxy clustering using these simulations are presented in \citet{McClintock2018} and \citet{Zhai2017} respectively. Additionally, we hope to provide convergence guidelines for future work that simulates the statistics presented here.

In \autoref{sec:params} we present our cosmological parameter space and the Latin Hypercube algorithm used to sample from it. In \autoref{sec:simulations} we discuss our simulation framework. In \autoref{sec:nbodyconv} we show that the observables we emulate are converged with respect to the choices made in our \nbody\ solver. In \autoref{sec:halofinding} we discuss issues related to halo finding and halo definitions, and in \autoref{sec:otheremu} we compare our simulations to existing emulators. In \autoref{sec:datarelease}, we discuss our plans to release these simulations to the public, and in \autoref{sec:summary} we conclude. 

\section{Cosmological Parameter Space}
\label{sec:params}

The goal of the parameter selection algorithm is to optimally span a large-dimensional space with a limited number of points. Our criterion for optimization is to maximize the accuracy of any scheme to interpolate statistics between the points, which requires the points to be as close to uniformly spaced as possible, while covering as much of the space as possible. We follow the technique outlined in \cite{Heitmann2009}, with minor modifications. The process begins with a Latin Hypercube (LH) containing $M=40$ samples of our $N=7$-dimensional space. In an LH design, each  of the $N$ dimensions is divided into $M$ bins. In each dimension, each of the bins is selected once with no repeats, thus guaranteeing the full range of each parameter within the space is represented sparsely.

\begin{figure*}
\includegraphics[width=\linewidth]{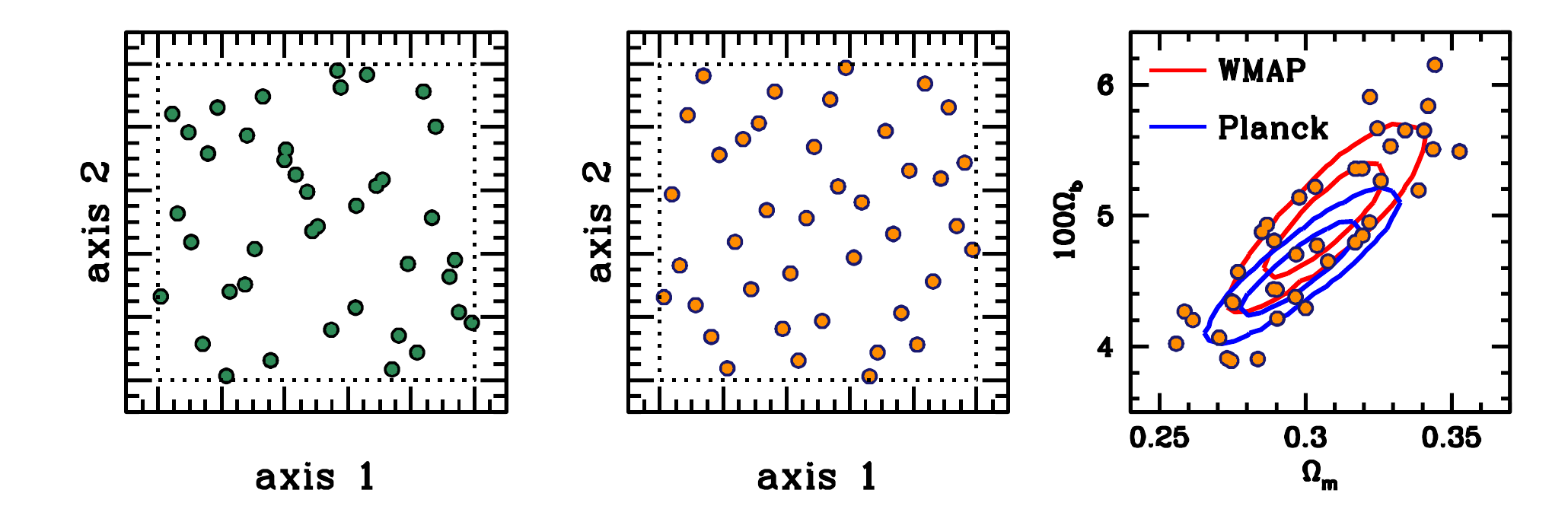}
\label{fig:hypercube}
\caption{{\it Left Panel:} A two-dimensional projects of a random 7-dimensional Latin Hypercube (LH), with 40 points in total. {\it Middle Panel:} The same LH, now optimized for more uniform spacing between points. {\it Right Panel:} The same LH as shown in the middle panel, but now rotated into the eigenspace defined by CMB data. Contours are the WMAP9 and Planck13 joint constraints with BAO and supernovae. 
}
\end{figure*}

A random LH design is not optimally spaced, however. Points can be clumped together, as shown in a two-dimensional projection of our 7-dimensional space in the left-hand side of  \autoref{fig:hypercube}. To quantify the spacing of a given LH, for every point in the space we calculate the distance to the closest point in each two-dimensional projection of the space. The quantity of interest is the sum of all minimum distances for all points in all projections. The space is optimal when this quantity is maximized, thus removing any clumping between points and pushing the points to a uniform distribution. To accomplish this, we use an iterative procedure that takes two points from the sample and swaps values in one dimension. If this swapping increases the quantity of interest, the swap is accepted. If it does not, the swap is rejected. This procedure is iterated until convergence. The result of this procedure is shown in the middle panel of  \autoref{fig:hypercube}. 

An LH design, by construction, creates a distribution of points in an $M$-dimensional cube. However, we do have prior knowledge on the distribution of cosmological parameters, and we want the distribution of our points to follow the degeneracies between parameters given current constraints. We use the combination of CMB, baryon acoustic oscillations (BAO), and supernovae (SN). Specifically, we use the CosmoMC chains produced in the cosmology analysis of the BOSS DR11 BAO analysis (\citealt{Anderson2014}). Separate chains were run for nine-year WMAP results \citep{WMAP9} and for Planck 2013 results \citep{Planck2013}. Given the differences in these CMB results, as well as our desire for our simulations to span a larger volume of parameter space than current constraints, we combine the chains from WMAP and Planck. The eigenvalues and eigenvectors of the combined chains are used to set the dimensions of the LH design. The generic LH design has 7 dimensions, with points ranging from [0,1]. Each of these dimensions is an eigenvector of the cosmological parameter space, and the range [0,1] maps onto $[-4,4]\times \sigma_i$, where $\sigma_i$ is the eigenvalue of vector $i$. The right-hand panel in \autoref{fig:hypercube} shows the generic LH design projected into cosmological parameter space. In this example, we plot $\Omega_m$ vs. $100\Omega_b$. In this projection, the data points may appear somewhat clumped, but recall that this is an angled projection of the LH. For reference, the $1\sigma$ and $2\sigma$ contours from the CMB+BAO+SN analyses are presented for both WMAP and Planck.

\begin{figure*}
\includegraphics[width=\linewidth]{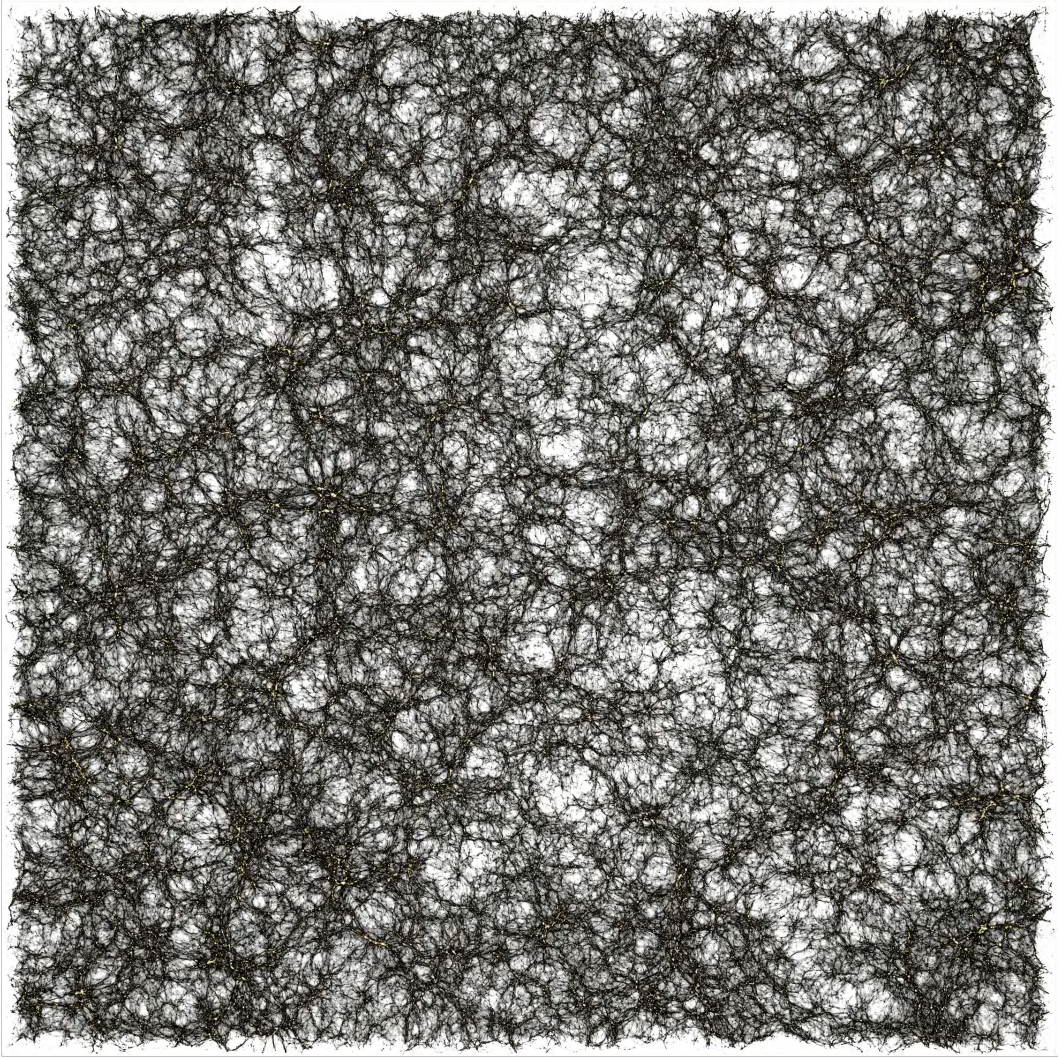}
\label{fig:slice}
\caption{A $50~\hmpc$ thick slice through \textsc{B25} with density deposition performed as described in \citet{Kaehler2012}.}
\end{figure*}

\begin{figure*}
\centering
\includegraphics[]{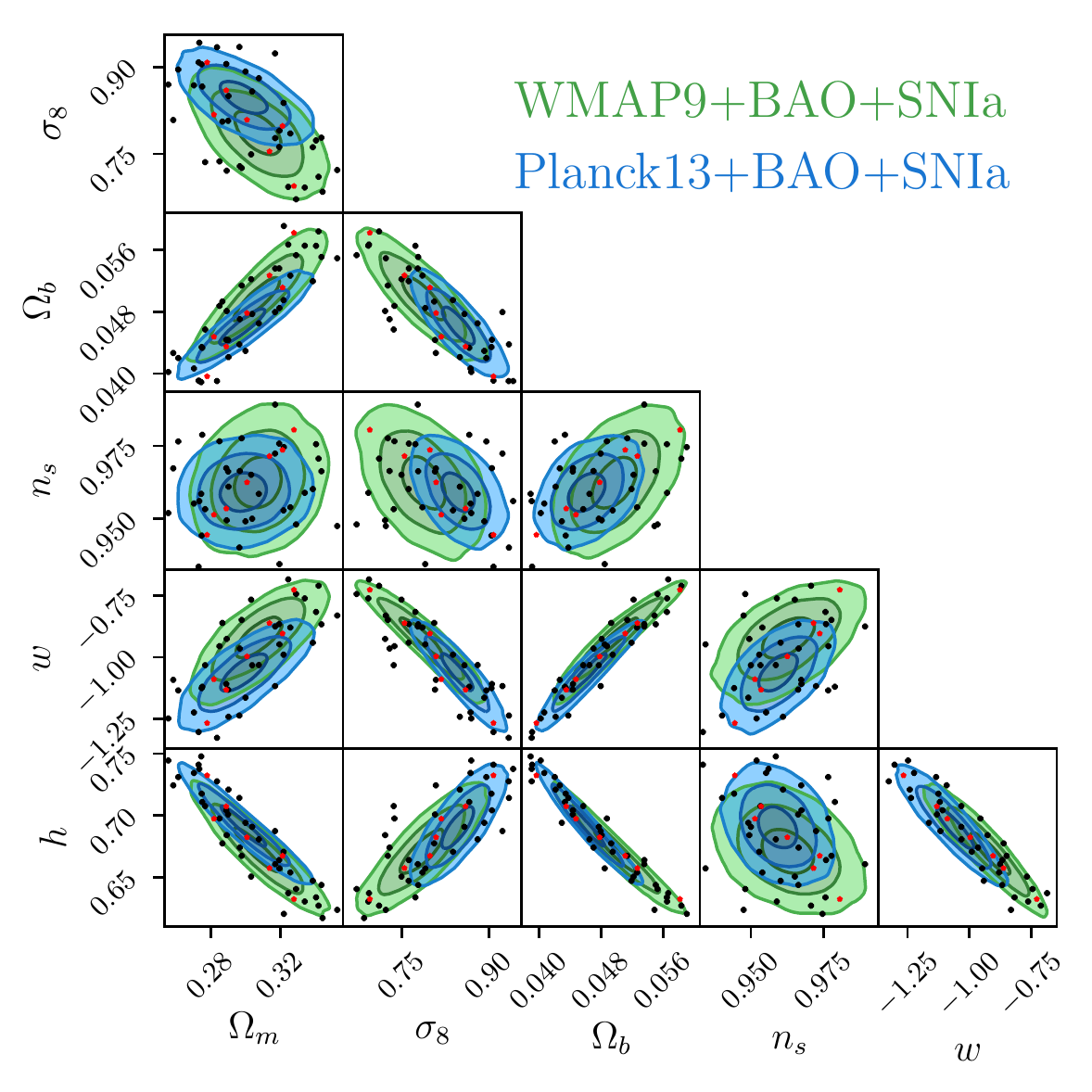}
\label{fig:allcontours}
\caption{The contours show the $3\sigma$ CMB+BAO+SNIa constraints in our parameter space. The 40 training cosmologies and seven test cosmologies are shown in black and red respectively.}
\end{figure*}

\begin{table*}[t!]
\centering
	\begin{tabular}{l|c|c|C|c|c|c|c}
	Name & $\Omega_{b}\/h^{2}$ & $\Omega_{c}\/h^{2}$ & $\textrm{w}_{0}$ & $n_{s}$ & $\log 10^{10}\textrm{A}_{s}$ & $\textrm{H}_{0}$ & $\textrm{N}_{\text{eff}}$\\
    \hline
B00 & 0.0227 & 0.1141 & -0.817 & 0.9756 & 3.093 & 63.37 & 2.919 \\ 
B01 & 0.0225 & 0.1173 & -1.134 & 0.9765 & 3.150 & 73.10 & 3.174 \\ 
B02 & 0.0230 & 0.1087 & -0.685 & 0.9974 & 3.094 & 63.71 & 3.259 \\ 
B03 & 0.0227 & 0.1123 & -0.744 & 0.9481 & 3.001 & 64.04 & 3.556 \\ 
B04 & 0.0221 & 0.1063 & -0.767 & 0.9651 & 3.119 & 65.05 & 2.664 \\ 
B05 & 0.0207 & 0.1295 & -1.326 & 0.9278 & 3.024 & 72.75 & 2.961 \\ 
B06 & 0.0229 & 0.1115 & -0.710 & 0.9706 & 3.016 & 62.70 & 2.706 \\ 
B07 & 0.0228 & 0.1196 & -0.867 & 0.9663 & 3.162 & 64.37 & 3.939 \\ 
B08 & 0.0207 & 0.1238 & -1.164 & 0.9491 & 3.147 & 69.40 & 3.599 \\ 
B09 & 0.0213 & 0.1158 & -0.831 & 0.9475 & 3.072 & 62.36 & 3.896 \\ 
B10 & 0.0219 & 0.1290 & -1.241 & 0.9610 & 3.050 & 72.09 & 4.236 \\ 
B11 & 0.0226 & 0.1090 & -0.861 & 0.9960 & 3.158 & 67.73 & 2.834 \\ 
B12 & 0.0225 & 0.1168 & -0.879 & 0.9540 & 3.048 & 65.38 & 2.876 \\ 
B13 & 0.0219 & 0.1172 & -1.120 & 0.9788 & 3.068 & 71.08 & 3.004 \\ 
B14 & 0.0226 & 0.1271 & -1.117 & 0.9724 & 3.094 & 68.73 & 2.749 \\ 
B15 & 0.0215 & 0.1285 & -1.303 & 0.9336 & 3.094 & 74.10 & 3.726 \\ 
B16 & 0.0218 & 0.1207 & -1.131 & 0.9662 & 3.014 & 70.07 & 3.769 \\ 
B17 & 0.0223 & 0.1194 & -1.248 & 0.9520 & 3.035 & 74.44 & 3.216 \\ 
B18 & 0.0229 & 0.1157 & -1.032 & 0.9533 & 3.020 & 70.75 & 4.279 \\ 
B19 & 0.0224 & 0.1133 & -1.092 & 0.9673 & 3.096 & 72.43 & 3.684 \\ 
B20 & 0.0223 & 0.1225 & -0.990 & 0.9529 & 3.120 & 67.06 & 3.386 \\ 
B21 & 0.0236 & 0.1172 & -0.866 & 0.9758 & 3.132 & 66.39 & 3.854 \\ 
B22 & 0.0215 & 0.1210 & -1.032 & 0.9586 & 3.072 & 68.06 & 2.621 \\ 
B23 & 0.0227 & 0.1012 & -0.566 & 0.9746 & 3.019 & 62.03 & 3.471 \\ 
B24 & 0.0225 & 0.1103 & -0.761 & 0.9589 & 3.144 & 63.03 & 4.151 \\ 
B25 & 0.0209 & 0.1171 & -0.948 & 0.9345 & 3.037 & 65.71 & 3.089 \\ 
B26 & 0.0224 & 0.1192 & -1.125 & 0.9443 & 3.128 & 71.76 & 2.791 \\ 
B27 & 0.0214 & 0.1134 & -0.965 & 0.9664 & 3.015 & 67.39 & 4.024 \\ 
B28 & 0.0217 & 0.1318 & -1.400 & 0.9586 & 3.147 & 74.77 & 3.811 \\ 
B29 & 0.0223 & 0.1289 & -1.236 & 0.9401 & 3.159 & 71.41 & 3.429 \\ 
B30 & 0.0219 & 0.1239 & -1.224 & 0.9552 & 3.118 & 73.43 & 4.066 \\ 
B31 & 0.0212 & 0.1276 & -1.382 & 0.9561 & 3.076 & 73.76 & 3.344 \\ 
B32 & 0.0225 & 0.1128 & -0.926 & 0.9495 & 3.043 & 68.40 & 3.981 \\ 
B33 & 0.0234 & 0.1150 & -0.875 & 0.9892 & 3.149 & 66.05 & 3.641 \\ 
B34 & 0.0228 & 0.1222 & -1.032 & 0.9500 & 3.107 & 69.07 & 3.131 \\ 
B35 & 0.0234 & 0.1076 & -0.613 & 0.9956 & 3.140 & 61.69 & 3.046 \\ 
B36 & 0.0220 & 0.1213 & -1.108 & 0.9674 & 3.179 & 70.41 & 3.301 \\ 
B37 & 0.0229 & 0.1097 & -0.849 & 0.9776 & 3.072 & 66.73 & 3.514 \\ 
B38 & 0.0237 & 0.1150 & -0.955 & 0.9766 & 3.054 & 69.75 & 4.109 \\ 
B39 & 0.0217 & 0.1201 & -0.941 & 0.9602 & 3.093 & 64.70 & 4.194 \\ 
	\end{tabular}
	\caption{The cosmologies used in training our emulators, deemed training cosmologies in this paper. Each has one realization with volume $(1050 \,\ \hmpc)^3$ and $N_{\text{part}}=1400^3$.  Each uses the fiducial settings detailed in \autoref{sec:simulations}. In particular they have mass resolutions of $3.51\times 10^{10} \left(\frac{\Omega_{m}}{0.3}\right) ~\hmsun$ and force resolutions of $20\,\ \hkpc$.}
	\label{table:trainingsims}
\end{table*}

\begin{table*}[t!]
\centering
	\begin{tabular}{l|c|c|C|c|c|c|c}
	Name & $\Omega_{b}\/h^{2}$ & $\Omega_{c}\/h^{2}$ & $\textrm{w}_{0}$ & $n_{s}$ & $\log 10^{10}\textrm{A}_{s}$ & $H_{0}$ & $N_{\text{eff}}$\\
    \hline
T00 & 0.0233 & 0.1078 & -0.727 & 0.9805 & 3.039 & 63.23 & 2.950 \\ 
T01 & 0.0228 & 0.1128 & -0.862 & 0.9715 & 3.064 & 65.73 & 3.200 \\ 
T02 & 0.0223 & 0.1178 & -0.997 & 0.9625 & 3.089 & 68.23 & 3.450 \\ 
T03 & 0.0218 & 0.1228 & -1.132 & 0.9535 & 3.114 & 70.73 & 3.700 \\ 
T04 & 0.0213 & 0.1278 & -1.267 & 0.9445 & 3.139 & 73.23 & 3.950 \\ 
T05 & 0.0218 & 0.1153 & -1.089 & 0.9514 & 3.119 & 69.73 & 3.700 \\ 
T06 & 0.0228 & 0.1203 & -0.904 & 0.9736 & 3.059 & 66.73 & 3.200 
	\end{tabular}
	\caption{The cosmologies used in the test simulations. Each has five realizations, each with volume $(1050\,\ \hmpc)^3$ and $N_{\text{part}}=1400^3$ using the fiducial settings detailed in \autoref{sec:simulations}. In particular, they have mass resolutions of $3.51\times 10^{10} \left(\frac{\Omega_{m}}{0.3}\right) ~\hmsun$ and force resolutions of $20\,\ \hkpc$.}
	\label{table:testsims}
\end{table*}

\section{\nbody\ Simulations}
\label{sec:simulations}
There are three sets of simulations discussed in this work, all run using the \textsc{L-Gadget2} \nbody\ solver, a version of \textsc{Gadget2} \citep{Springel2005} modified for memory efficiency when running dark-matter-only (DMO) simulations. The first of these sets, which we dub ``training simulations'', is a set of 40 $(1.05~h^{-1} \/\textrm{Gpc})^{3}$ boxes with $1400^3$ particles, resulting in a mass resolution of $3.51\times 10^{10} \left(\frac{\Omega_{m}}{0.3}\right) ~\hmsun$. The cosmologies of these simulations, listed in \autoref{table:trainingsims}, are drawn from the LH discussed in \autoref{sec:params}. These are run with a Plummer equivalent force softening of $20~h^{-1}\/\textrm{kpc}$, and maximum time step of $\textsc{max}(\Delta \ln a)=0.025$. We use 2nd order Lagrangian perturbation theory (2LPT) initial conditions generated at $a=0.02$ using \textsc{2LPTIC} \citep{Crocce2006} with input power spectra as computed by CAMB \citep{CAMB}, taking $\Omega_{\nu}=0$. 

Each of these 40 simulations is initialized with a different random seed. This is different from the approach taken in some recent simulation suites designed for emulators \citep[e.g][]{Garrison2017}, but \citet{McClintock2018} and \citet{Zhai2017} show that this enables our emulators to perform better than the sample variance of our individual simulations, whereas simulations using the same initial seed are guaranteed to perform only as well as the sample variance of the chosen individual simulation volume. We save 10 snapshots at redshifts of $z=\{3.0, 2.0, 1.0, 0.85, 0.7, 0.55, 0.4, 0.25, 0.1, 0.0\}$. 

In order to test the accuracy of our emulators we have also run a set of seven test cosmologies using the same settings as our training simulations. For each test cosmology we have run 5 simulations, each with different initial conditions, totaling 35 simulations. We will refer to these as ``test simulations'' throughout. 

Additionally, we have run a set of simulations varying a number of choices with respect to the \textsc{L-Gadget2} \nbody\ solver, which we will refer to as ``convergence test simulations''. A few of these simulations were run using a number of cosmologies, including the Chinchilla cosmology \citep{Lehmann2017} with $(\Omega_m,h,N_{\text{eff}},n_s,\sigma_8,w) = (0.286\/,0.7\/,3.04,\/0.96,\/0.82,\/-1)$, which is not used for any of the test or training boxes, but is well within our cosmological parameter space. See \autoref{table:convsims} for a summary of the simulations that we have used for these tests.

The names of these simulations all begin with \textsc{CT} to denote that they were run for convergence testing. The first number following \textsc{CT} enumerates the \nbody\ solver parameter set that was used to run the simulation. Various sets of these simulations were run with the same random seed for their initial conditions. The sets with the same seed are demarcated with the same last number in their name, e.g. \textsc{CT00} and \textsc{CT60}. When necessary, we distinguish between the different cosmologies used by including them in the simulation name. For example, \textsc{CT00-T00} refers to the simulation run with our fiducial \nbody\ solver parameters using the first random seed for its initial conditions in the \textsc{T00} cosmology as listed in \autoref{table:testsims}. 

We have chosen to use volumes of $(400~\hmpc)^3$ for the \textsc{CT} simulations rather than the $(1.05~\hgpc)^3$ used for the training and test simulations, as changing some of the settings for convergence tests significantly increases the runtime of the simulations. Using a smaller volume allows these simulations to complete in a modest amount of time. We have mitigated the smaller volumes by running 4 pairs of boxes, \textsc{CT00,...,CT03} and \textsc{CT40,...,CT43}, in 3 different cosmologies, \textsc{Chinchilla}, \textsc{T00} and \textsc{T04}, for our particle loading test as it is for this test that we find our largest deviations from convergence and we wish to constrain these more precisely with better statistics.

We employ the \textsc{rockstar} spherical overdensity halo finder \citep{Behroozi2013} for all of our simulations. \textsc{rockstar} employs a 6D phase space friends-of-friends (FoF) algorithm in order to identify density peaks and their surrounding overdensities. We have chosen to use $\mathrm{M}_\text{200b}$ strict spherical overdensity (SO) masses as our fiducial mass definition, where strict refers to the inclusion of unbound particles in the mass estimates of all halos. A discussion of this choice is presented in \autoref{sec:halofinding}. Other than enabling strict SO masses, we have used the default \textsc{rockstar} settings, choosing the \textsc{rockstar} softening length to be the same as that used in the \nbody\ solver and leaving on the particle downsampling that \textsc{rockstar} performs in its initial construction of friends-of-friends groups, as we find that this does not affect any of the conclusions presented in this work. Additionally, all results presented here use only host halos, i.e. halos which are not found to lie within a halo with a higher maximum circular velocity. 

\section{$N$-Body Convergence Tests}
\label{sec:nbodyconv}
The sampling of the parameter space, the effective volume of the training set, and the details of the emulators are all important aspects in determining the final precision and accuracy of our predictions. Equally important is that the observables that are used to train the emulators are converged with respect to all possible choices made when running the simulations. Otherwise, there is a risk of biasing the predictions in ways that are difficult to identify post-hoc. For instance, comparison with other predictions is a useful sanity check so long as they agree to within their purported precision as it is unlikely that both sets of simulations have the same systematic biases, and so their agreement indicates that both predictions are likely converged. However, in the case that such comparisons disagree it is impossible to determine why unless detailed convergence tests are conducted.

It should be noted again that all of the tests we perform are of relative convergence and not absolute convergence. The reasons for this are twofold. First, we are knowingly leaving out physics that we believe to be important at some level, such as the effects of baryonic feedback on the matter distribution. Additionally, we do not have an analytic solution towards which we are measuring convergence even for the physics that we have implemented, particularly in the non-linear regime. There is a growing literature on the possible lack of absolute convergence in \nbody\ simulations in this regime \citep{vandenbosch2018a,vandenbosch2018b}, but these issues typically arise when considering dark matter substructure within host halos. Constraining the statistics of substructure is not the goal of the present work, and so we conduct no tests of convergence of such statistics here.  

\begin{table*}[htbp!]
\centering
	\begin{tabular}{l|c|c|c|c|c|c|c|c|c}
	Name & Cosmology & $N_{realizations}$ & $L_{box}$ $[h^{-1}\textrm{Mpc}]$ & $m_{\text{part}}$ $[h^{-1}\textrm{M}_{\odot}]$& $\epsilon$ $[h^{-1} \textrm{kpc}]$& $\Delta \ln a_{max}$ & $a_{\text{start}}$ & $\alpha$& $\eta$\\
    \hline
	\textsc{CT0} & Chinchilla, T00, T04 & $3\times4$ & $400$ & $3.30\times 10^{10}\left(\frac{\Omega_{m}}{0.286}\right)$ & 20 & 0.0250 & 0.02 & 0.002& 0.0250\\
  	\textsc{CT1} & Chinchilla & 1 & $400$ & $3.30\times 10^{10}$ & 20 & 0.0250 & 0.01 & 0.002& 0.0250\\
    \textsc{CT2} & Chinchilla & 1 & $400$ & $3.30\times 10^{10}$ & 10 & 0.0250 & 0.02 & 0.002& 0.0250\\
	\textsc{CT3} & Chinchilla & 1 & $400$ & $3.30\times 10^{10}$ & 20 & 0.0250 & 0.02 & 0.001& 0.0250\\
	\textsc{CT4} & Chinchilla & 1 & $400$ & $3.30\times 10^{10}$ & 20 & 0.0125 & 0.02 & 0.002& 0.0250\\
	\textsc{CT5} & Chinchilla & 1 & $400$ & $3.30\times 10^{10}$ & 20 & 0.0250 & 0.02 & 0.002& 0.0125\\ 
    \textsc{CT6} & Chinchilla, T00, T04 & $3\times4$ & $400$ & $4.12\times 10^{9}\left(\frac{\Omega_{m}}{0.286}\right)$ & 20 & 0.0250 & 0.02 & 0.002& 0.0250 \\
    \textsc{CT7} & T00,...,T06 & 7 & $3000$ & $2.49\times 10^{12} \left(\frac{\Omega_{m}}{0.286}\right)$ & 20 & 0.0250 & 0.02 & 0.002& 0.0250\\

	\end{tabular}
	\caption{Summary of the boxes run for convergence tests. Columns are simulation name, cosmologies, number of different initial condition realizations, box side length, particle mass, Plummer equivalent force softening, maximum time step, starting scale factor, force error tolerance, and time integration error tolerance.}
	\label{table:convsims}
\end{table*}

\subsection{Measurements}
\label{subsec:measurements}
Below we describe our measurements of the following observables:

\begin{enumerate}[label=\alph*)]
\item Matter power spectrum, $P(k)$,
\item 3-dimensional matter correlation function, $\xi_{mm}(r)$,
\item Spherical overdensity halo mass function, $N(M_\text{200b})$
\item 3-dimensional halo--halo correlation function, $\xi_{hh}(r)$,
\item Projected galaxy--galaxy correlation function, $w_{p}(r_{p})$, and
\item Monopole and quadrupole moments of the redshift space galaxy--galaxy correlation function, $\xi_0(s)$, $\xi_{2}(s)$.
\end{enumerate}

We briefly detail how we measure each of these statistics and describe our convergence tests in the following subsections. 
\subsubsection{Matter Power Spectrum}
The first statistic we will be interested in is the matter power spectrum, $P(k)$, which is given by
\begin{align}
\langle \delta(\mathbf{k}) \delta(\mathbf{k^{\prime}})^{*} \rangle = \delta_{\mathbf{k},\mathbf{k^{\prime}}} P(k)
\end{align}
where $\delta(\mathbf{k})$ is the Fourier transform of the matter overdensity field $\delta(\mathbf{x}) = \frac{\rho(\mathbf{x}) - \bar{\rho}}{\bar{\rho}}$:
\begin{align}
\delta(\mathbf{x}) = \frac{1}{(2\pi)^{3}}\int d\mathbf{k}~ e^{-i\mathbf{k\cdot x}}~\delta(\mathbf{k})
\end{align}
and the angle brackets denote an ensemble average over independent volumes, $V$. The power spectrum fully describes the statistics of any Gaussian random field, and as such is a useful statistic for describing cosmological density fields, which are still Gaussian at most scales until late times.

Because our simulations have periodic boundary conditions, we can estimate the power spectrum using a Fast Fourier Transform (FFT). First, we deposit the density field onto a mesh of dimensions $N_{mesh}^3$ where $N_{mesh}=\frac{2 L_{box} k_{max}}{\pi}$ using a cloud-in-cell deposition, such that wavenumbers $k\le k_{max}$ are sampled at or above their Nyquist rate. We take $k_{max}=5 ~\invhmpc$. We then compensate for the mass-deposition window function and average the resulting $3D$ power spectrum in bins of $k$, with $dk = \frac{L_{box}}{2\pi}$. All of this is performed using the \textsc{python} package \textsc{nbodykit} \citep{nbodykit}. We do not perform any shot-noise subtraction, since for the scales we are using, the standard $\frac{V}{N_{\text{part}}}$ correction is small, and it is not clear that the correction should necessarily take this form. Unless otherwise noted, this is the only statistic for which we do not include error estimates. At the scales of interest, the errors on $P(k)$ are very small, and estimating them via jackknife as we have done for our other measurements is non-trivial due to our use of an FFT to measure $P(k)$.

\subsubsection{$3D$ Matter Correlation Function}
Since $\delta(\mathbf{x})$ is assumed to be a stationary random field, its correlation function is given by the Fourier transform of its power spectrum,
\begin{align}
\xi(r) &= \langle \delta(\mathbf{x}) \delta(\mathbf{x} + \mathbf{r}) \rangle \\
	   &= \frac{1}{(2\pi)^{3}} \int d\mathbf{k} ~P(k)~e^{-i\mathbf{k}\cdot \mathbf{r}}
\end{align}

We estimate the $3D$ matter correlation function from our boxes by jackknifing the Landy-Szalay estimator \citep{Landy93}
\begin{align}
\label{eq:LS}
\hat{\xi}(r) &= \frac{DD - 2DR + RR}{RR}
\end{align}
where $DD$, $DR$ and $RR$ are particle-particle, particle-random, and random-random pair counts normalized by the total number of possible pairs in a given radial bin. We use 27 jackknife regions and down-sample the particle distribution by a factor of 100 which we have checked does not affect our results.

Despite the simple relation between the matter power spectrum and $3D$ correlation function, we check the convergence of both since measurement errors take different forms in the two statistics, e.g. in configuration space, correlations functions are formally only affected by shot-noise at $r=0$, whereas for power spectra the correction affects all wavenumbers. Additionally, emulators built from these boxes may choose to use one or the other quantity, and determining the scales or wavenumbers where one of these statistics is converged using the other is non-trivial. All pair counting was done using \textsc{Corrfunc} \citep{corrfunc}.

\subsubsection{Spherical Overdensity Halo Mass Function}
In modern theories of $\Lambda$CDM galaxy formation, all galaxies are assumed to form within dark matter halos. As such, making converged predictions for the abundance of dark matter halos is of great importance for accurately predicting galaxy statistics. In particular, \citet{McClintock2018} uses the simulations presented here to build an emulator for the abundance of SO dark matter halos using $\Delta=\text{200b}$ and so we focus our convergence tests on the statistic used in that work, namely the total number of halos per bin in $\log_{10}(M_\text{200b})$, $N(M_\text{200b})$. Additionally, we are interested in only so-called host halos and not subhalos. This is because the galaxy models we employ (in e.g. \citealt{Zhai2017}) are based on the Halo Occupation Distribution (HOD) formalism, which has no need for subhalo information, and because the first applications of our halo mass function emulator will be cosmology constraints using cluster number counts. We estimate $N(M_\text{200b})$ and its errors in our simulations using a jackknife estimator with 27 jackknife regions. 

\subsubsection{3D Halo Correlation Function}
The other diagnostic we will use to asses the convergence of our halo populations is the $3D$ halo correlation function, $\xi_{hh}(r)$. Because the clustering of halos is biased with respect to the matter distribution due to their preferential formation in overdense regions of the matter distribution, convergence of $\xi_{hh}$ at a particular scale, $r$, does not directly follow from convergence of $\xi_{mm}$ at the same scale, and thus it is important to test for convergence of these separately. 

For a discrete field, the two-point correlation function measures the excess probability, relative to a Poisson distribution, of finding two halos at the volume elements $dV_{1}$ and $dV_{2}$ separated by a distance $r$ \citep{Peebles1980}:
\begin{equation}
dP_{12} = \bar{n}^2[1+\xi(r)]dV_{1}dV_{2},
\end{equation}
where $\bar{n}$ is the mean number density of the sample. To estimate $\xi_{hh}$ we again jackknife the Landy-Szalay estimator given by \autoref{eq:LS}, using 27 subvolumes. The measurements of $\xi_{hh}(r)$ presented here are for halos with $M_\text{200b} > 10^{12}~\hmsun$, except where otherwise noted.

\subsubsection{Galaxy Correlation Functions}
In order to calculate galaxy clustering, we employ 10 BOSS massive galaxy sample-like HOD models and populate halos with a galaxy number density of $4.2\times10^{-4}\/(\hmpc)^{-3}$ at $z=0.55$.
The typical mass scale $M_{\rm{min}}$ for the dark matter halo at which half of the halos have a central galaxy is in the range $12.9 < \log{M_{\rm{min}}[h^{-1}{\rm M}_{\odot}]} < 13.5$, and the scatter of halo mass at fixed galaxy luminosity $\sigma_{\log{M}}$ ranges from 0.05 to 0.5. These models have satellite fraction ranging from $10\sim13\%$ and galaxy bias from $2.0\sim2.13$.  Satellites are assumed to have a NFW profile \citep{NFW} and their velocity dispersion is assumed to be independent from the position within the halo. More details of this HOD model can be found in \citet{Zhai2017}.

The correlation function of the resulting galaxy catalogs is described by the projected correlation function $w_{p}$ and redshift space multipoles $\xi_{l}$. The former is used to mitigate redshift space distortions, probing the real space clustering signal, while the later can be calculated via Legendre decomposition at given order $l$. In the calculation of $w_{p}$, the integral along the line of sight direction is truncated at 80 $\hmpc$, which is large enough to include most of the correlated pairs and produce a stable result. For $\xi_{l}$, we perform the decomposition up to $l=2$ to get the redshift space monopole and quadrupole. Clustering measurements are presented at scales from 0.1 to 50 $\hmpc$ and averaged over 10 random realizations of each of the 10 HOD models. 

\subsection{Convergence Tests}
Having described the statistics for which we will check convergence, we now report on the tests that we have performed as well as their results.
\subsubsection{Initial Conditions}
Since our \nbody\ simulations do not start at $a=0$ we must justify our choice of initial conditions. Two decisions must be made: 1) which analytic prescription to use to generate the initial density and velocity fields and 2) what epoch to generate the initial conditions. For the first, we use 2LPT. For the second, care must be taken to ensure that the analytic treatment used to generate the initial conditions remains accurate for all modes in the simulation until the scale factor at which we start the \nbody\ solver. To ensure this, we have chosen a starting time of $a=0.02$ ($z=49$). In this section we show that our observables are robust to this choice by comparing measurements made in a fiducial simulation with the same specifications used in our emulator suite, \textsc{CT00-Chinchilla}, to a simulation that has been run with a starting scale factor of $a=0.01$ ($z=99$), \textsc{CT10}. It should be noted again that, unless otherwise stated, all convergence tests presented here vary one parameter at a time from the fiducial parameters used for our training and test simulations.

\begin{figure*}[htbp!]
\begin{tabular}{ccc}
\centering
  \includegraphics[width=0.3\linewidth]{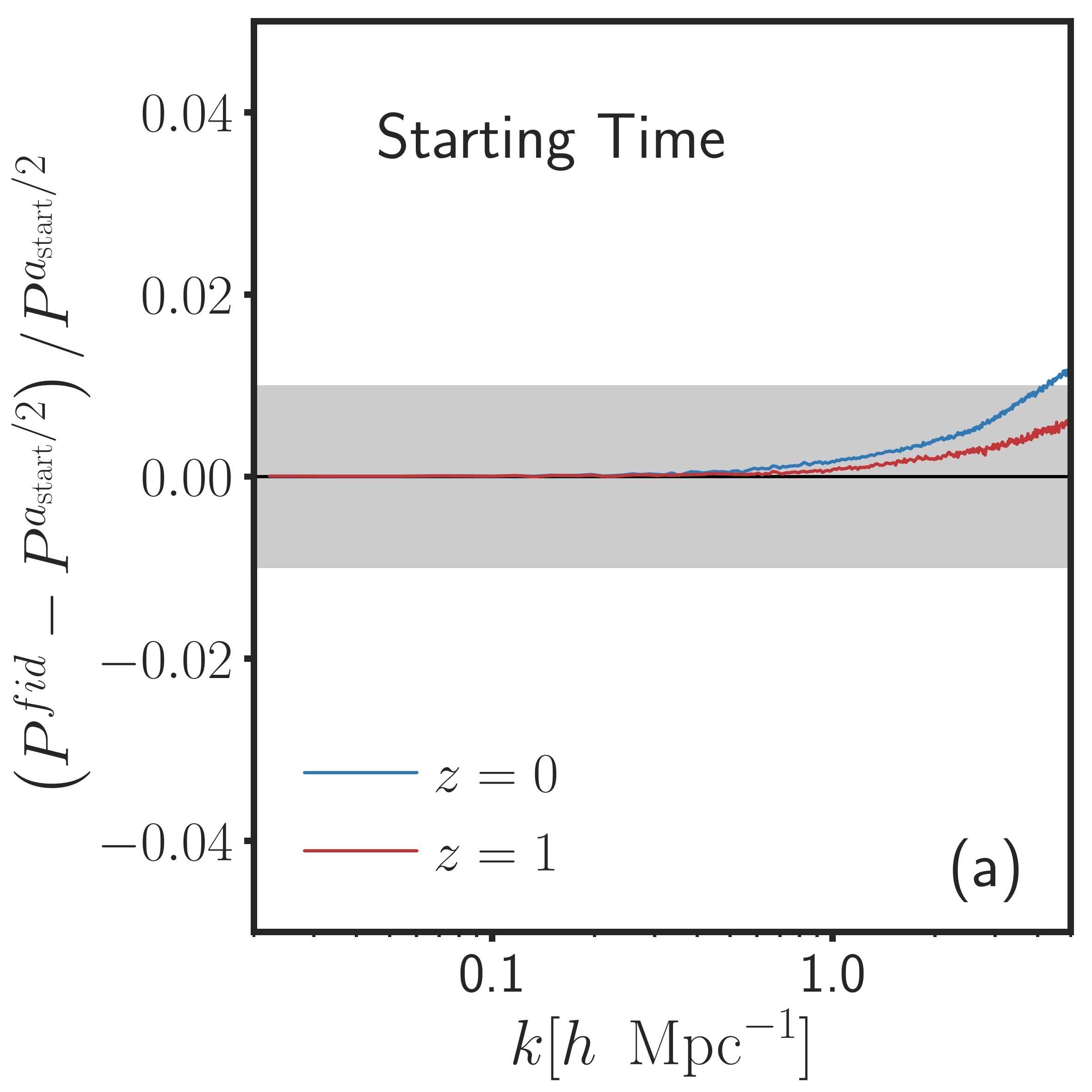} &
  \includegraphics[width=0.3\linewidth]{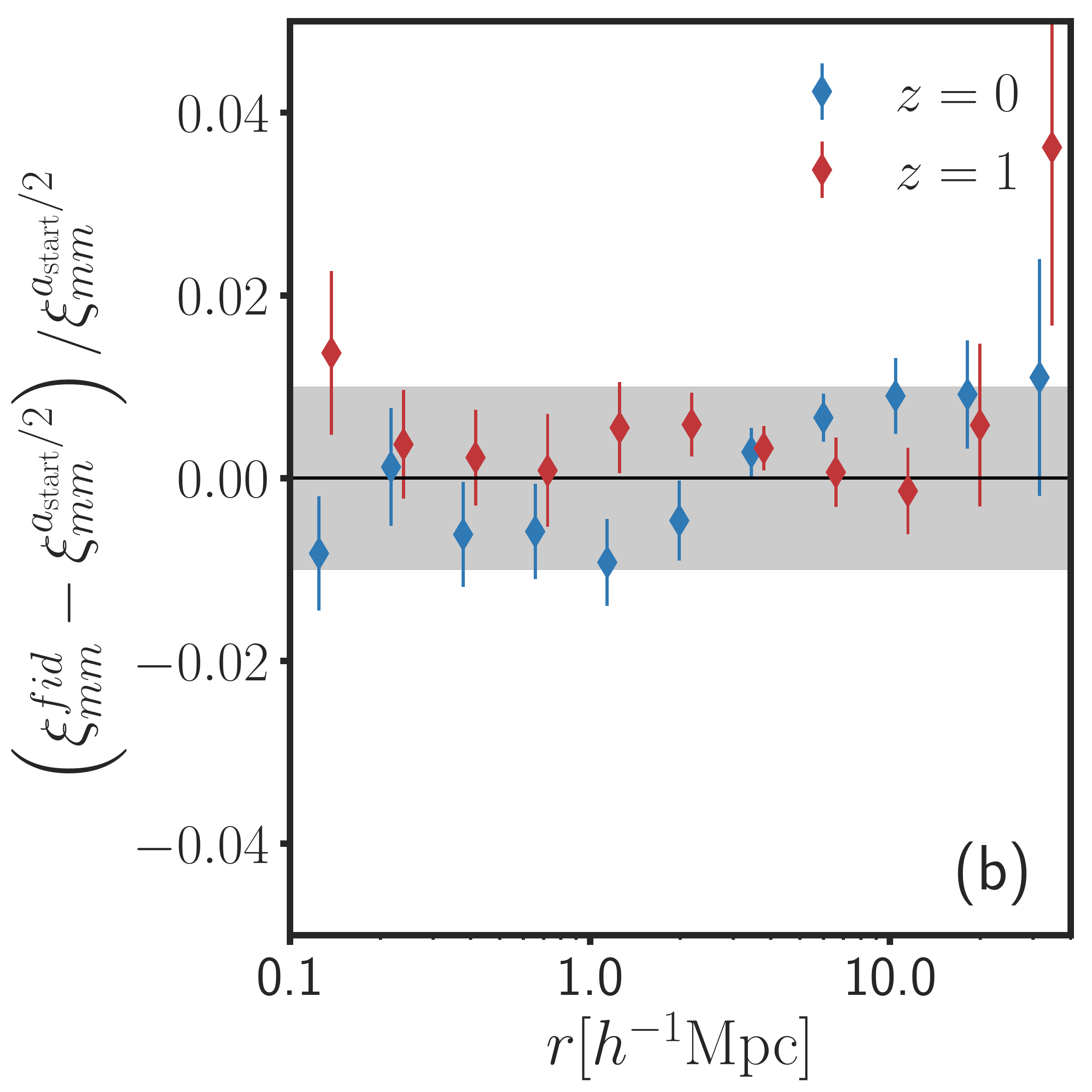} &
  \includegraphics[width=0.3\linewidth]{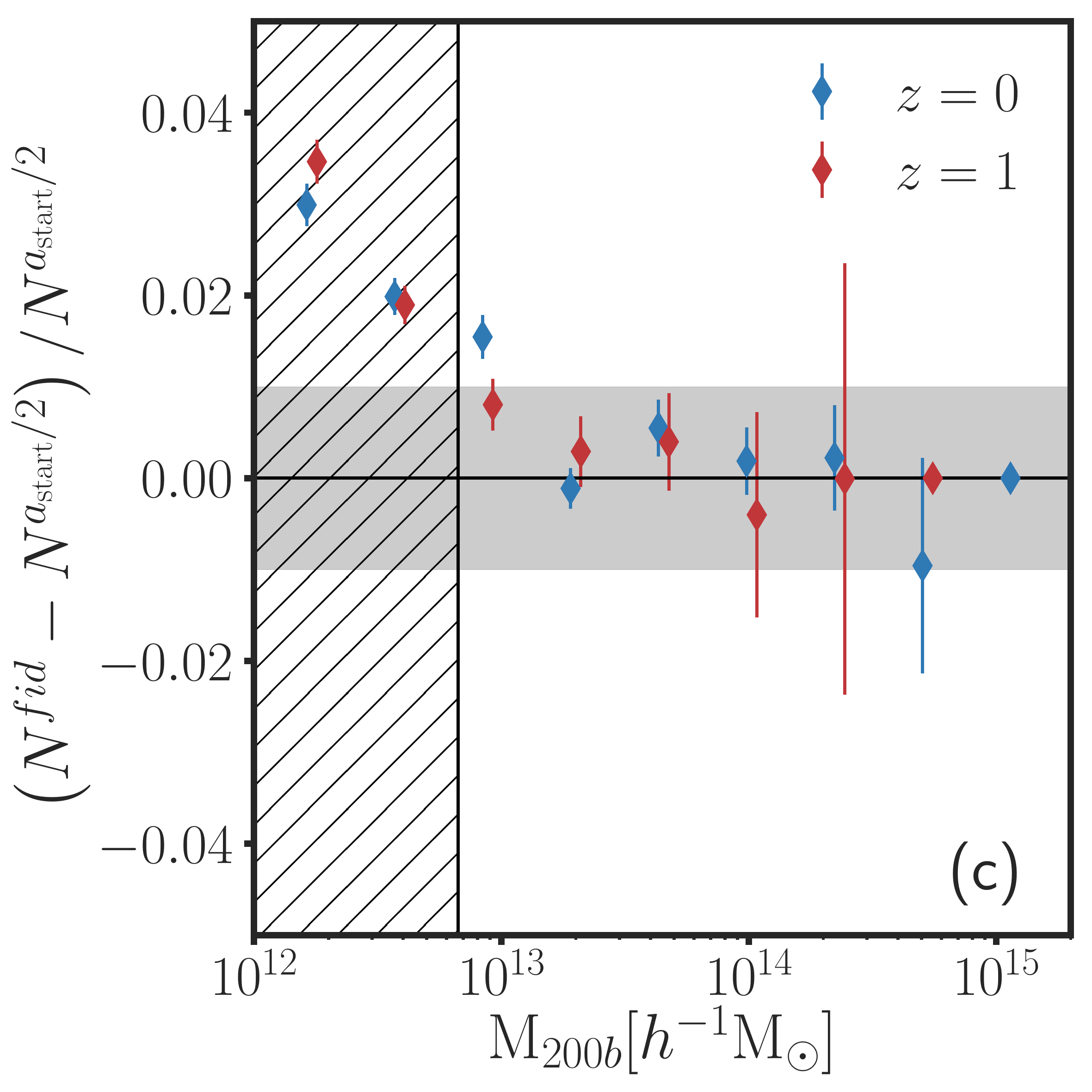} \\  
  \includegraphics[width=0.3\linewidth]{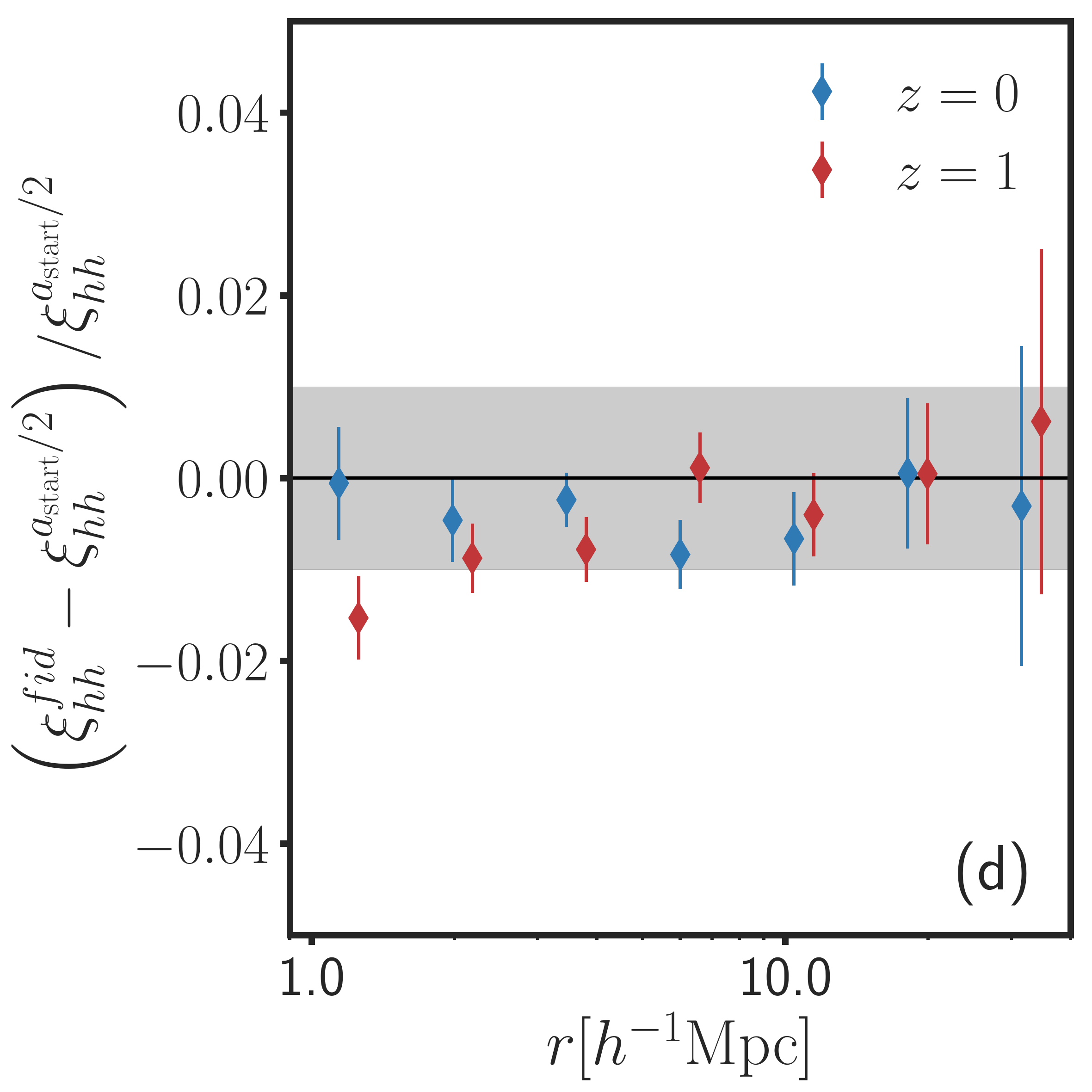}& 
 \includegraphics[width=0.3\linewidth]{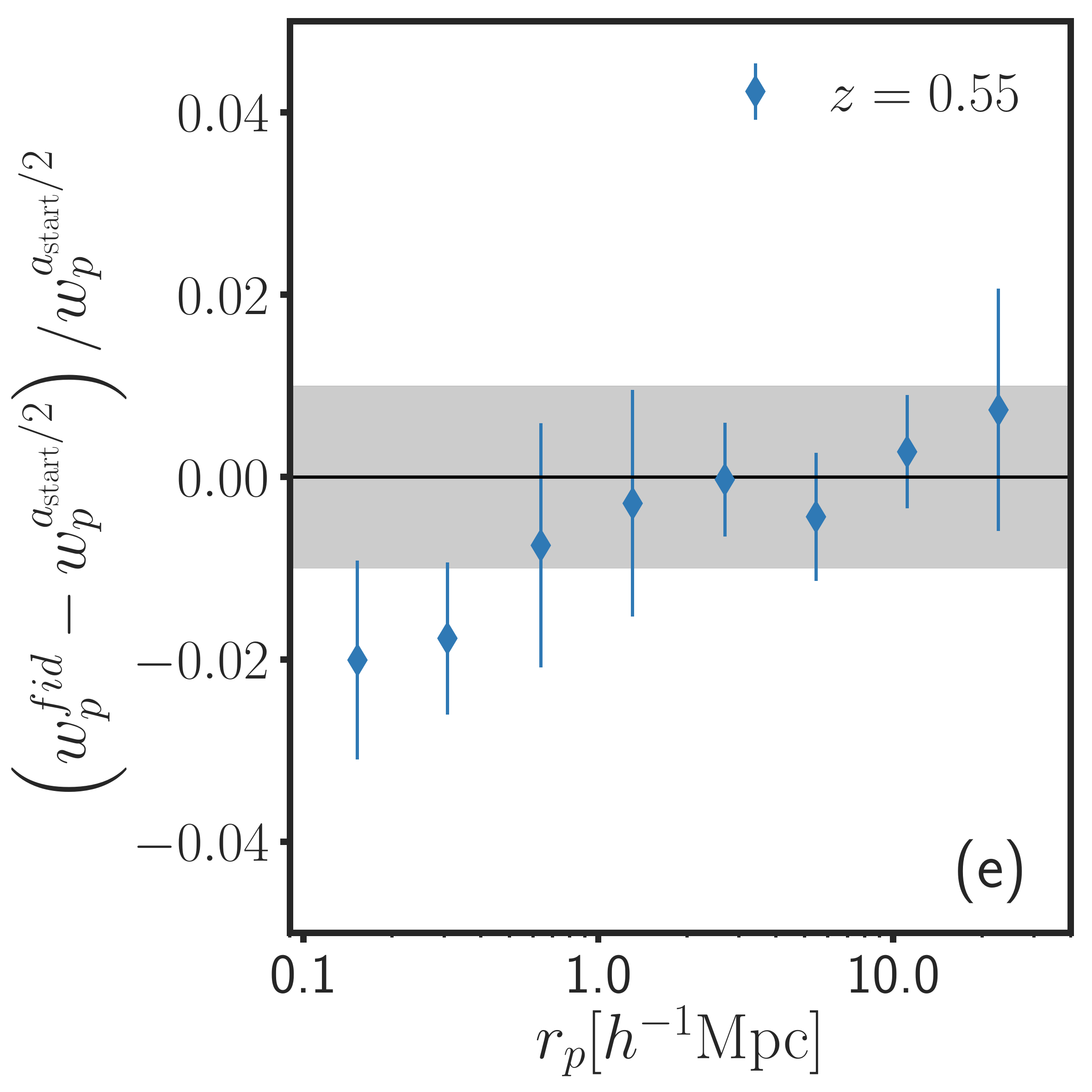} & \includegraphics[width=0.3\linewidth]{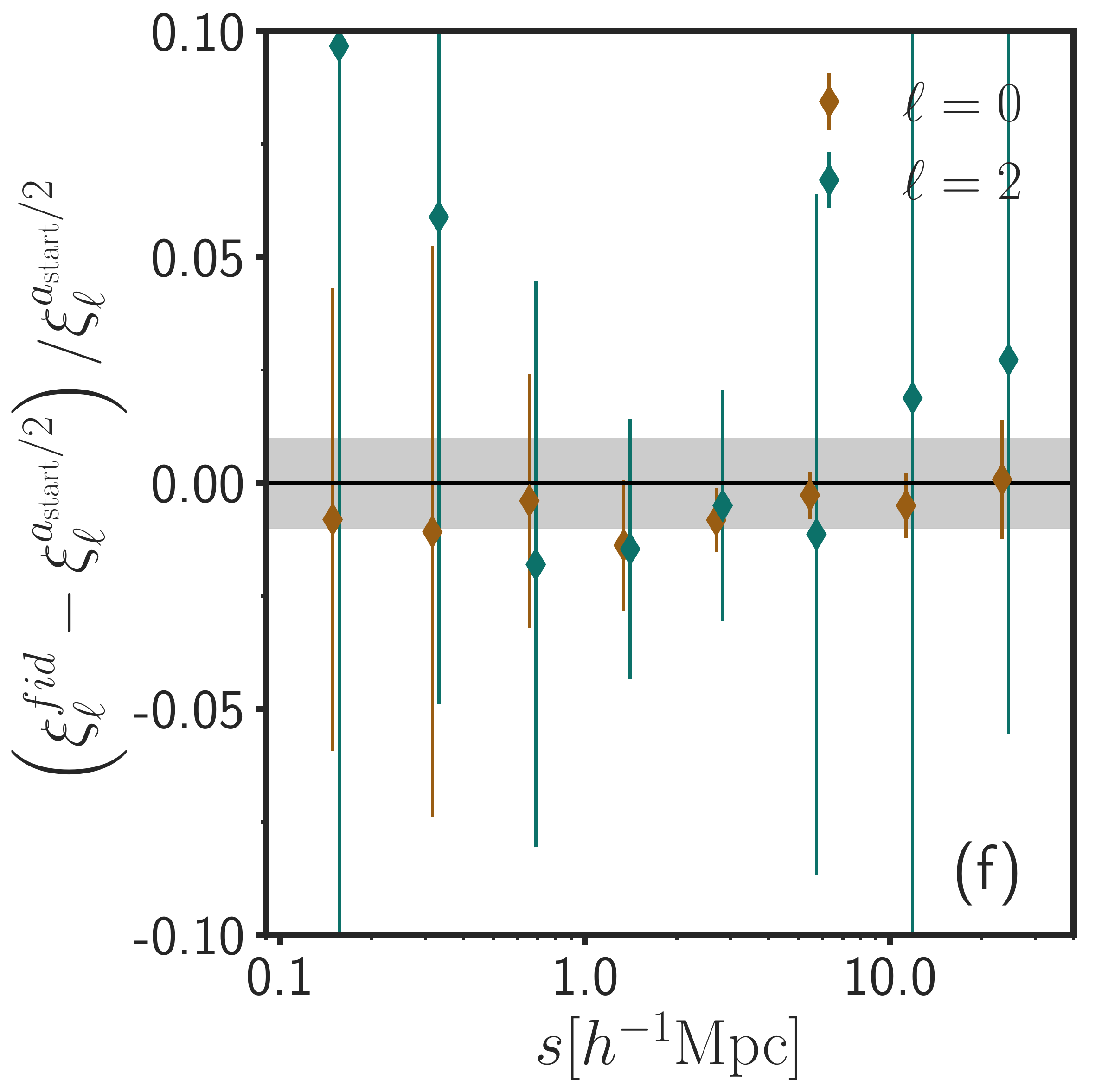} 
 \\

\end{tabular}
\caption{Comparison of a number of observables from \textsc{CT00-Chinchilla}, a simulation with our fiducial starting scale factor, $a=0.02$, and \textsc{CT10}, a simulation with a starting scale factor of $a=0.01$. The gray band in all figures denotes $1\%$ accuracy. \textbf{(a)} Matter power spectrum. \textbf{(b)} Matter correlation function. \textbf{(c)} Halo mass function, where the hatched region corresponds to halos with fewer than 200 particles. \textbf{(d)} Halo--halo correlation function for $M_\text{200b}>10^{12}~\hmsun$. \textbf{(e)} Galaxy projected correlation function averaged over 10 realizations of 10 different HODs. \textbf{(f)} Redshift space monopole and quadrupole for the same HODs.}
\label{fig:starttime}
\end{figure*}

For this test and all following tests, we will only report on deviations from $1\%$ convergence which exceed $1\sigma$ in significance. For this test, a few statistics deviate from convergence by more than this as can be seen in  \autoref{fig:starttime}. For $M_\text{200b}<10^{13} \hmsun$, the halo mass function deviates from convergence. This mass would fall in the lowest mass bin used in \cite{McClintock2018}, and is still within the range of halo masses used in HOD models in \cite{Zhai2017}. We also find deviations from convergence approaching $1\sigma$ in $w_{p}$ for $\sim r<300 \hkpc$, $P(k)$ for $k\sim 4~\invhmpc$ at $z=0$, and $\xi_{hh}(r)$ for $r \le 1~ \hmpc$. The largest scale data point of $\xi_{mm}(r)$ also deviates from $1\%$ convergence for $z=1$, but this is likely a statistical fluctuation given the convergence of $P(k)$ at large scales.

A likely explanation for the observed deviations from convergence is inaccuracies in 2LPT at these scales in describing non-linear evolution between $0.01<a<0.02$. If this is the case then these effects may also vary with cosmology such that cosmologies with less (more) structure growth at early times will deviate from convergence by less (more), and cosmologies with less (more) late time structure growth will have less (more) redshift evolution of this effect.

\subsubsection{Force Softening}
When solving the $N$-body problem as an approximation to collisionless dynamics, one must employ a so-called force softening in order to mitigate the effects of unphysical two body interactions. In \textsc{l-gadget2} this is done by representing the single particle density distribution as a Dirac delta function convolved with a spline kernel \citep{Monaghan1985} $\delta(\textbf{x}) = W(|\textbf{x}|, 2.8\epsilon)$, where $W(r)$ is given by a cubic spline.

Using this kernel, the potential of a point mass at $r=0$ for non-periodic boundary conditions is given by $-Gm/\epsilon$. It is this $\epsilon$ that we refer to as the force softening length. Typically, smaller $\epsilon$ yields equations which are closer to those that govern the true universe, but decreasing $\epsilon$ by too much at fixed mass resolution will lead to undesirable two body interactions as mentioned above. There is an extensive literature on convergence of various quantities with respect to force softening length \citep[e.g.]{Power2003}, but for completeness we investigate this convergence in the context of the exact statistics that we plan to measure and emulate with this simulation suite. 

\begin{figure*}[htbp!]
\centering
\begin{tabular}{ccc}
  \includegraphics[width=0.3\linewidth]{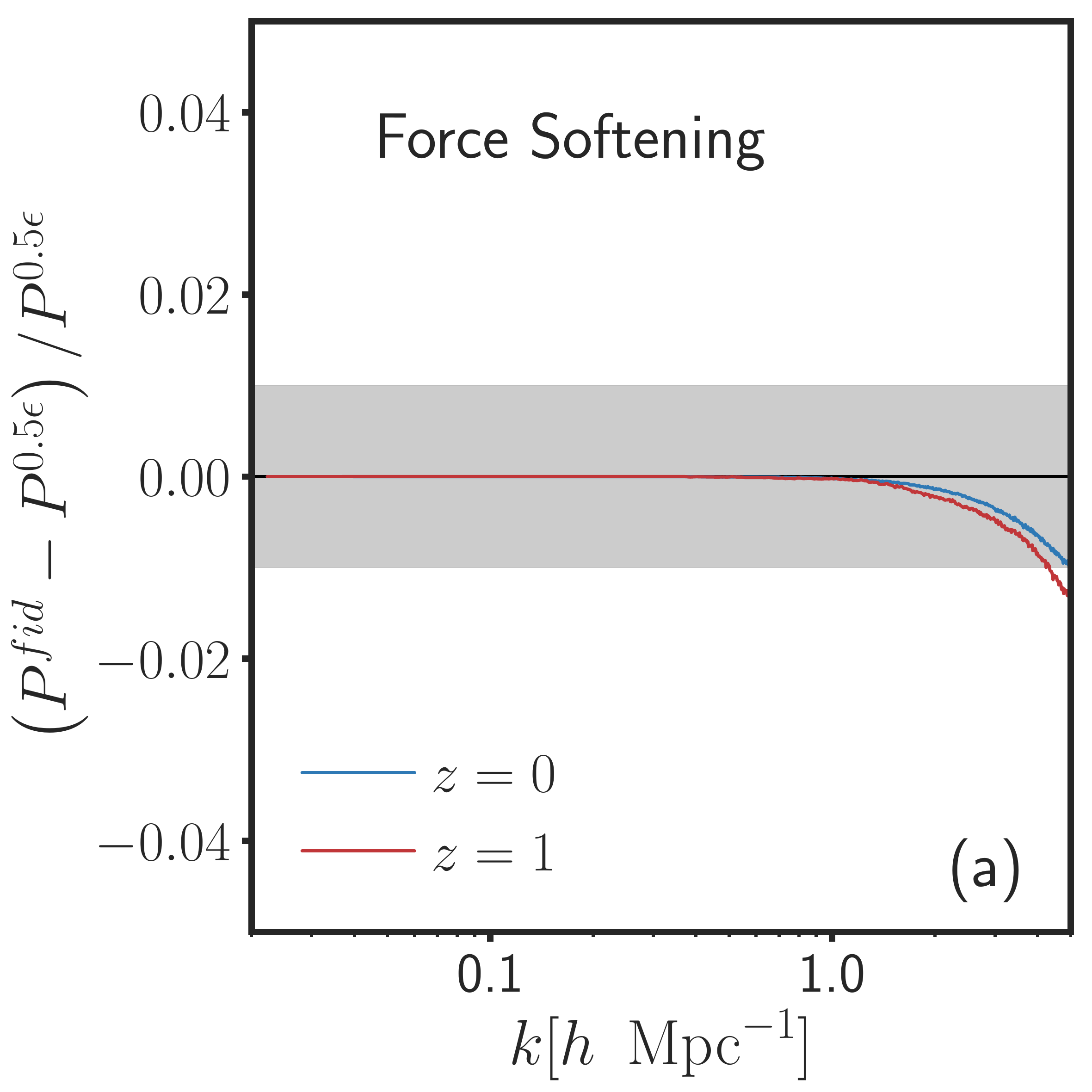} &
  \includegraphics[width=0.3\linewidth]{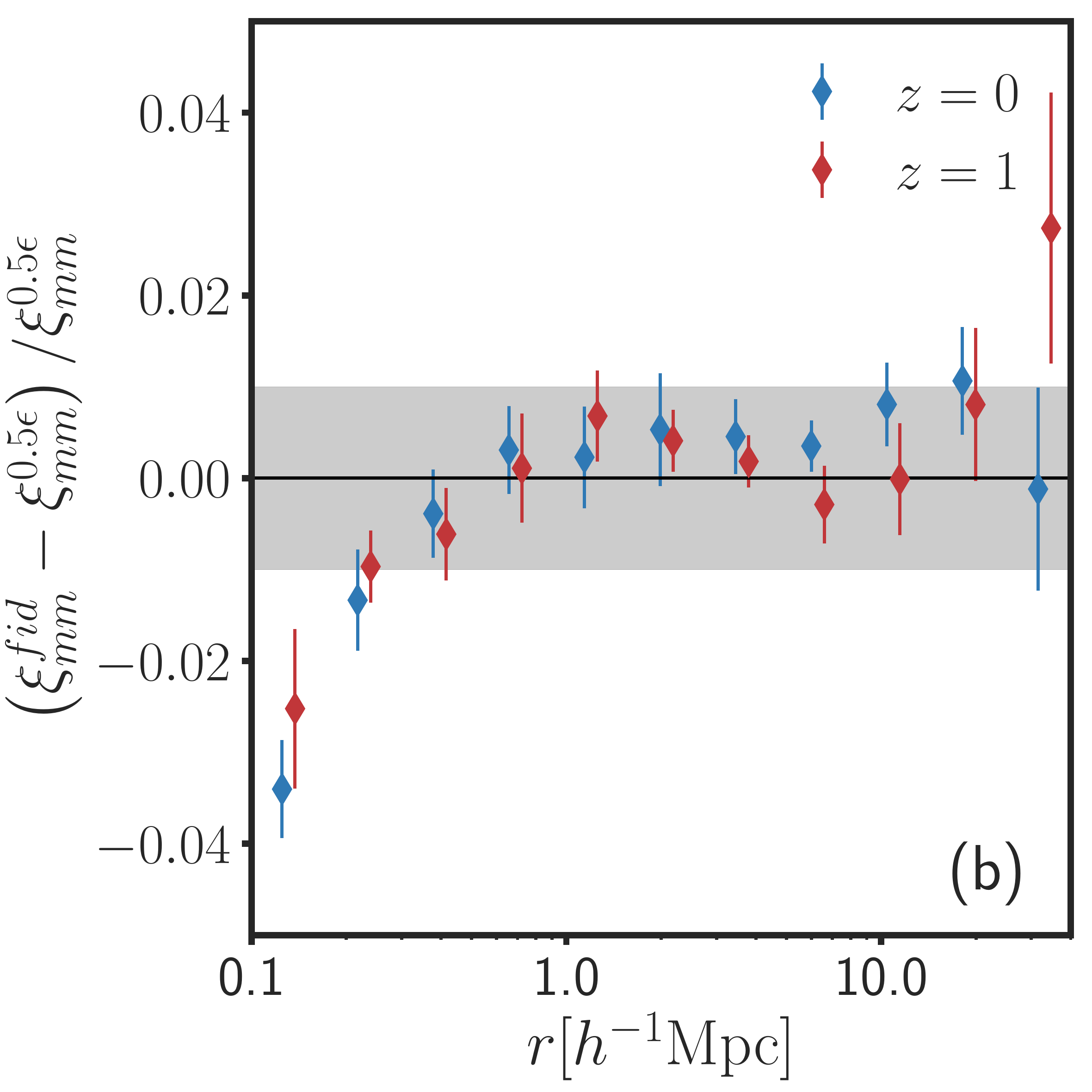} &  
   \includegraphics[width=0.3\linewidth]{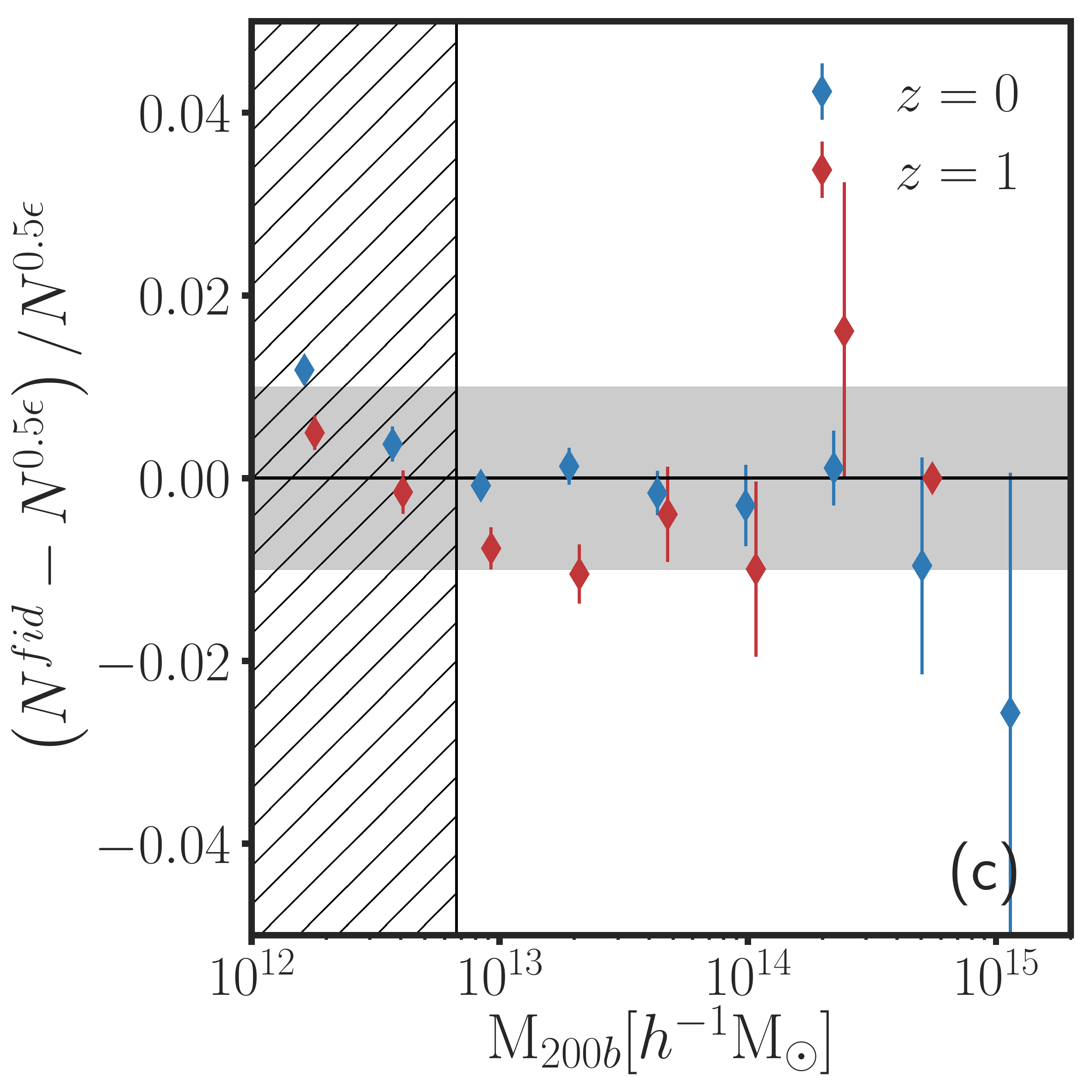} \\ 
   \includegraphics[width=0.3\linewidth]{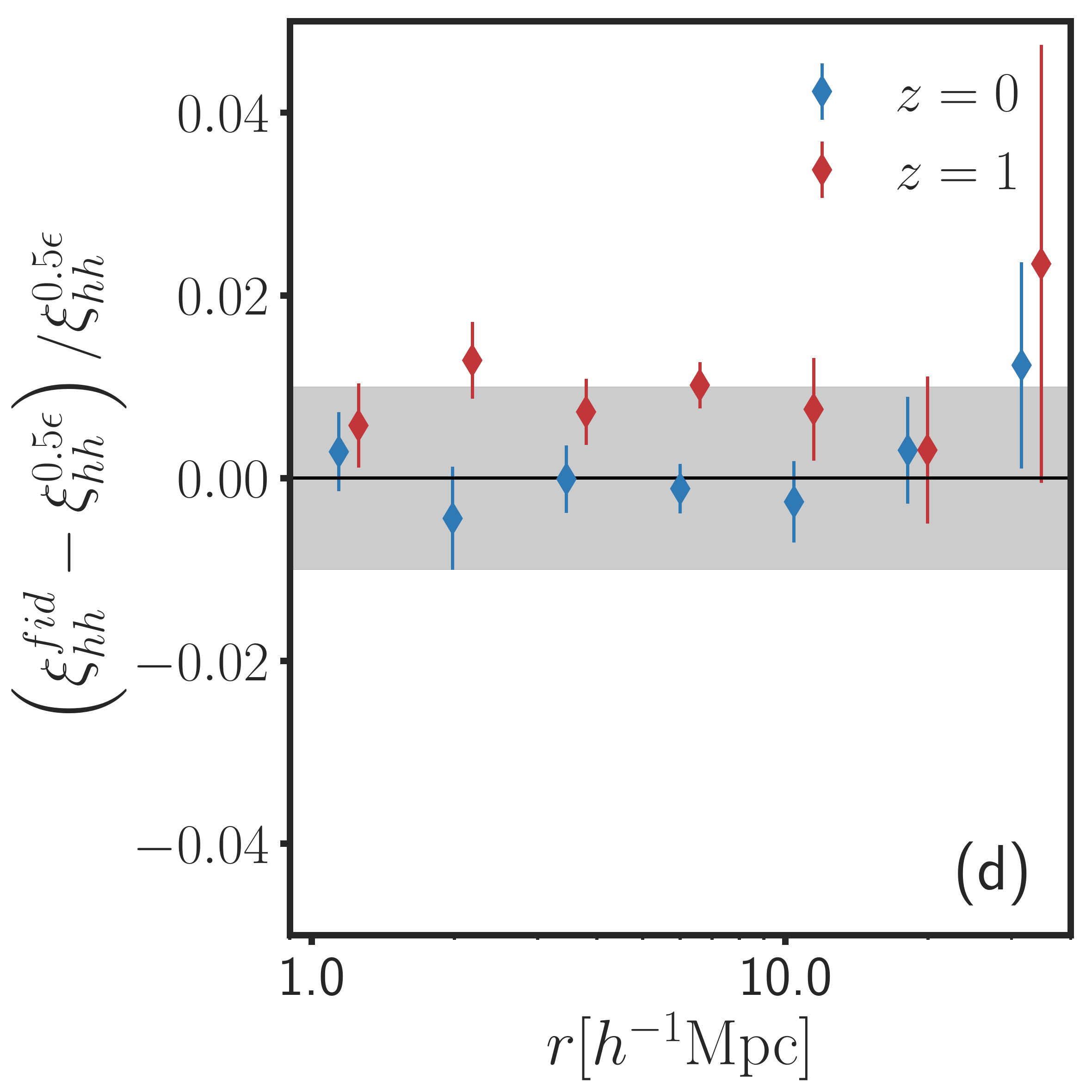}  & 
   \includegraphics[width=0.3\linewidth]{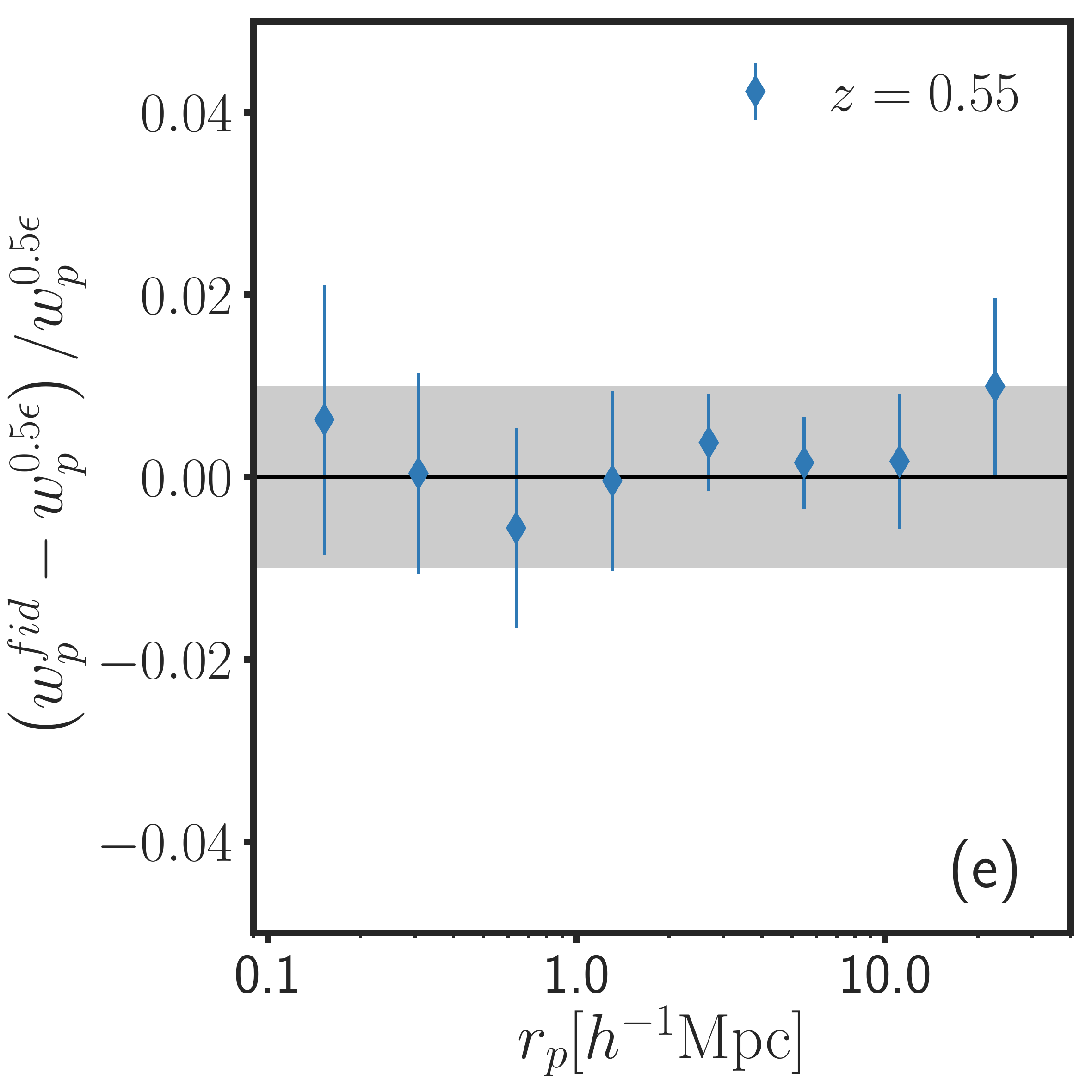} &
  \includegraphics[width=0.3\linewidth]{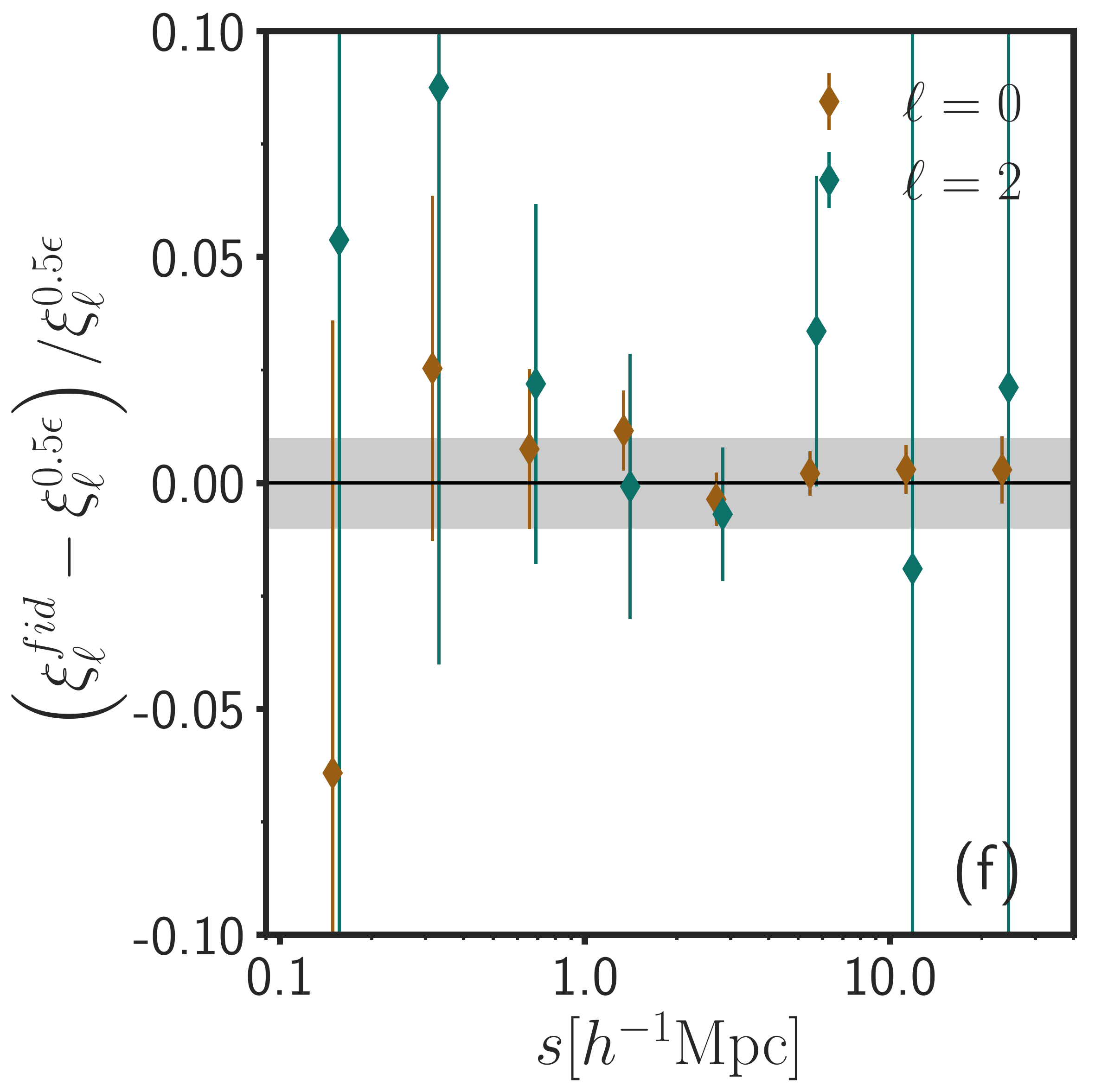} \\   
\end{tabular}
\caption{Convergence tests with respect to force softening. Observables measured from a simulation with our fiducial parameters, \textsc{CT00-Chinchilla} are compared to \textsc{CT20}, a simulation with half the force softening: $\epsilon=10\hkpc$.  Subfigures are the same as in \autoref{fig:starttime}.}
\label{fig:forcesoftening}
\end{figure*}

For our fiducial simulations, we have set $\epsilon=20~\hkpc$, and for this convergence test we have run an additional simulation, \textrm{CT20} with $\epsilon=10~\hkpc$. The results of the comparison between our fiducial simulation, \textrm{CT00-Chinchilla}, and \textrm{CT20} can be found in \autoref{fig:forcesoftening}. The only statistic which deviates from convergence is $\xi_{mm}(r)$ for $r<200~ \hkpc$. $P(k)$ is converged to the $1\%$ level for the scales measured here, but by $k\sim 3 ~\invhmpc$ is showing systematic deviations from perfect convergence. This is consistent with the findings of \citet{Heitmann2010}, although they only consider $\epsilon>25~\hkpc$. The deviations seen in $\xi_{mm}$ and $P(k)$ do not appear to have a significant effect on other statistics.

\subsubsection{Absolute Force Error Tolerance}
Another parameter which governs the accuracy of the gravitational force calculations is how deeply to walk the octtree used to partition space when summing the small-scale contributions to the gravitational force on each particle. This is typically referred to as the cell opening criterion, since it is used to determine whether or not a cell in the tree should be ``opened'' and traversed. We use the standard \textsc{l-gadget2} relative opening criterion which opens a cell containing mass $M$, extension $l$ at a distance from the point under consideration of $r$ if 

\begin{align}
\frac{GM^2}{r^2}\left(\frac{l}{r}\right)^{2} > \alpha |\textbf{a}_{\rm{old}}|
\end{align}
where $|\textbf{a}_{\rm{old}}|$ is the magnitude of the acceleration of the particle under consideration in the last time step and $\alpha$ is a free parameter allowing tuning of the accuracy. In general, decreasing $\alpha$ leads to smaller errors in force computation, but greater run time as more nodes in the tree must be opened per time step. Our fiducial runs use $\alpha=0.002$. 

\begin{figure*}[htbp!]
\begin{tabular}{ccc}
\centering
  \includegraphics[width=0.3\linewidth]{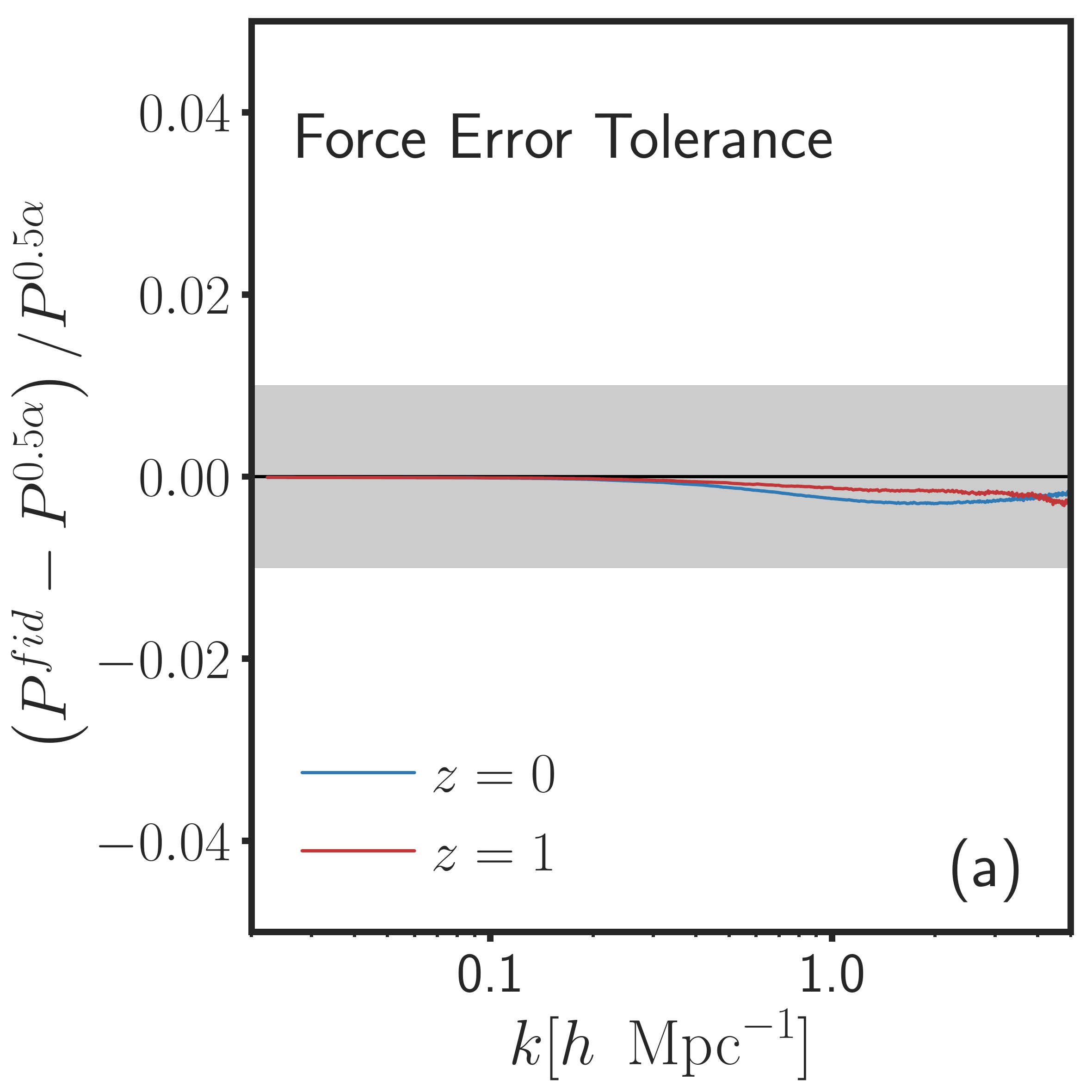} &
\includegraphics[width=0.3\linewidth]{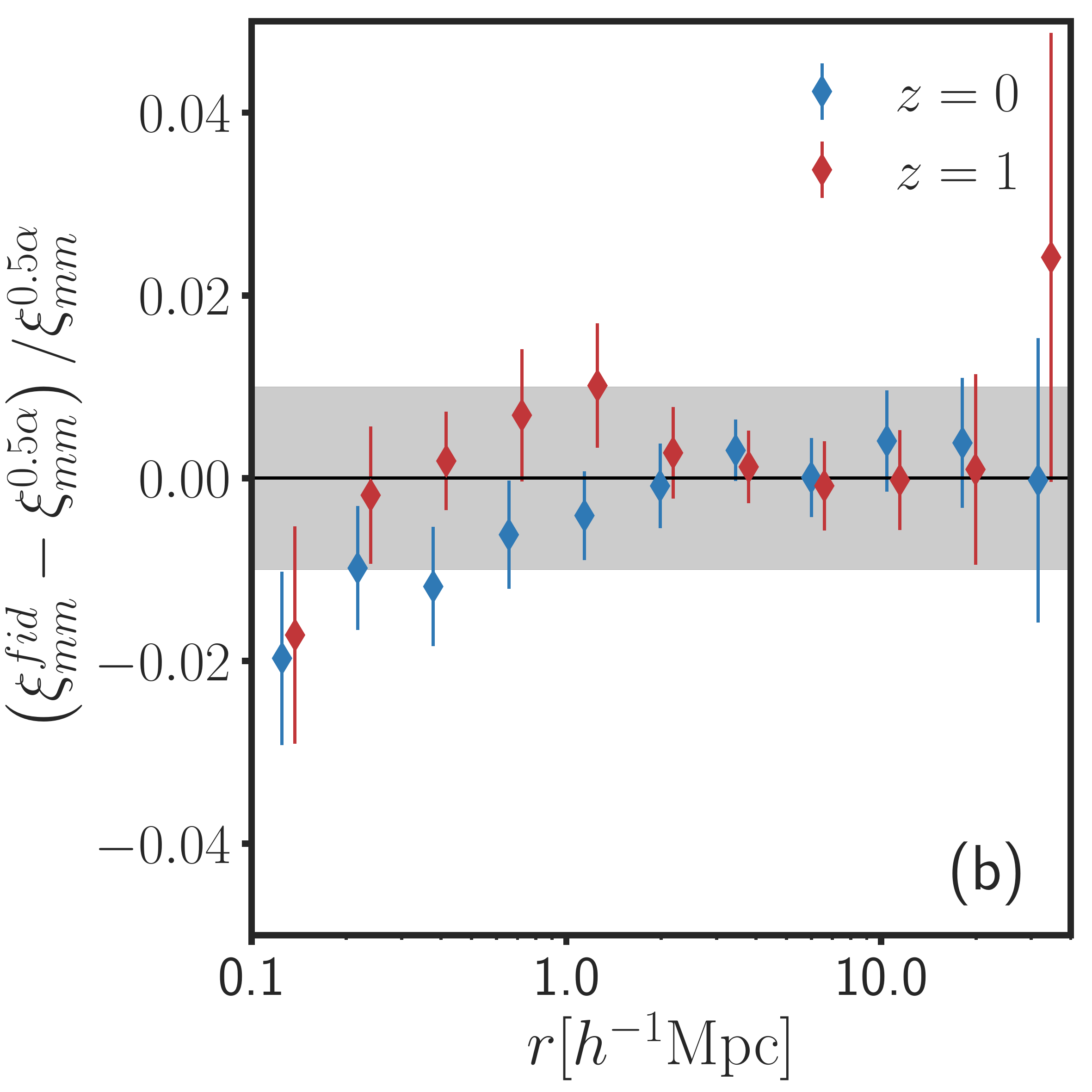} &
  \includegraphics[width=0.3\linewidth]{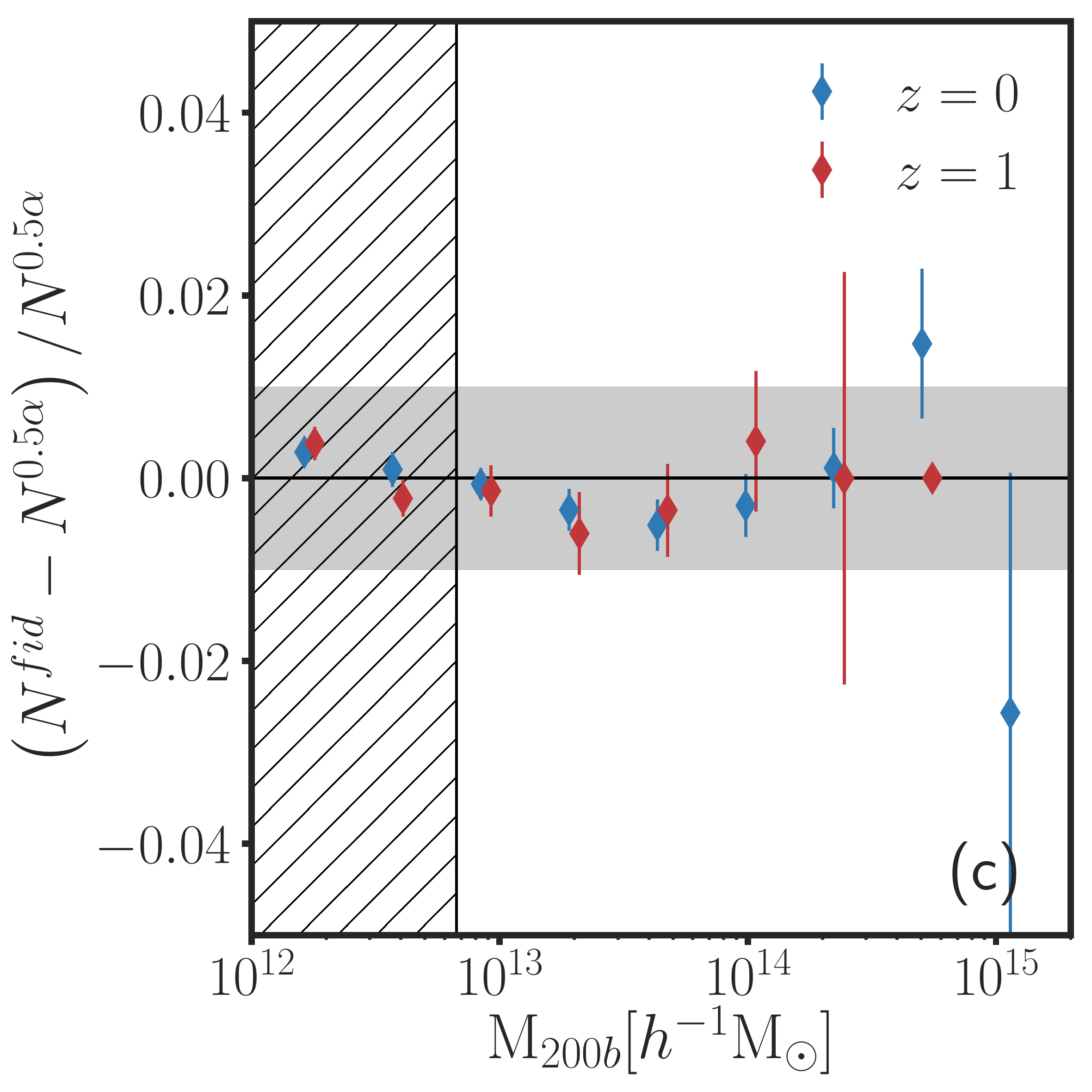}\\ 
  \includegraphics[width=0.3\linewidth]{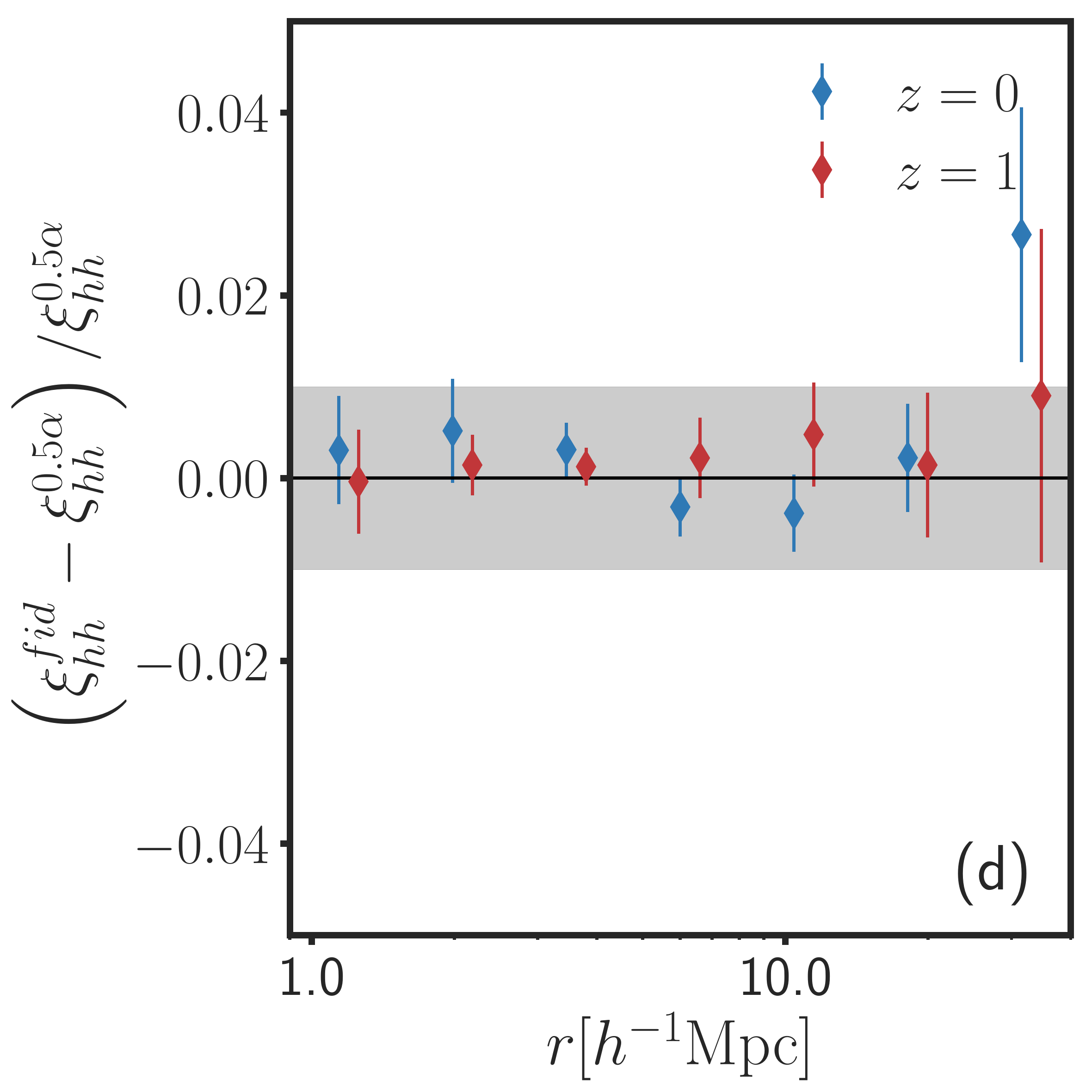}  & 
  \includegraphics[width=0.3\linewidth]{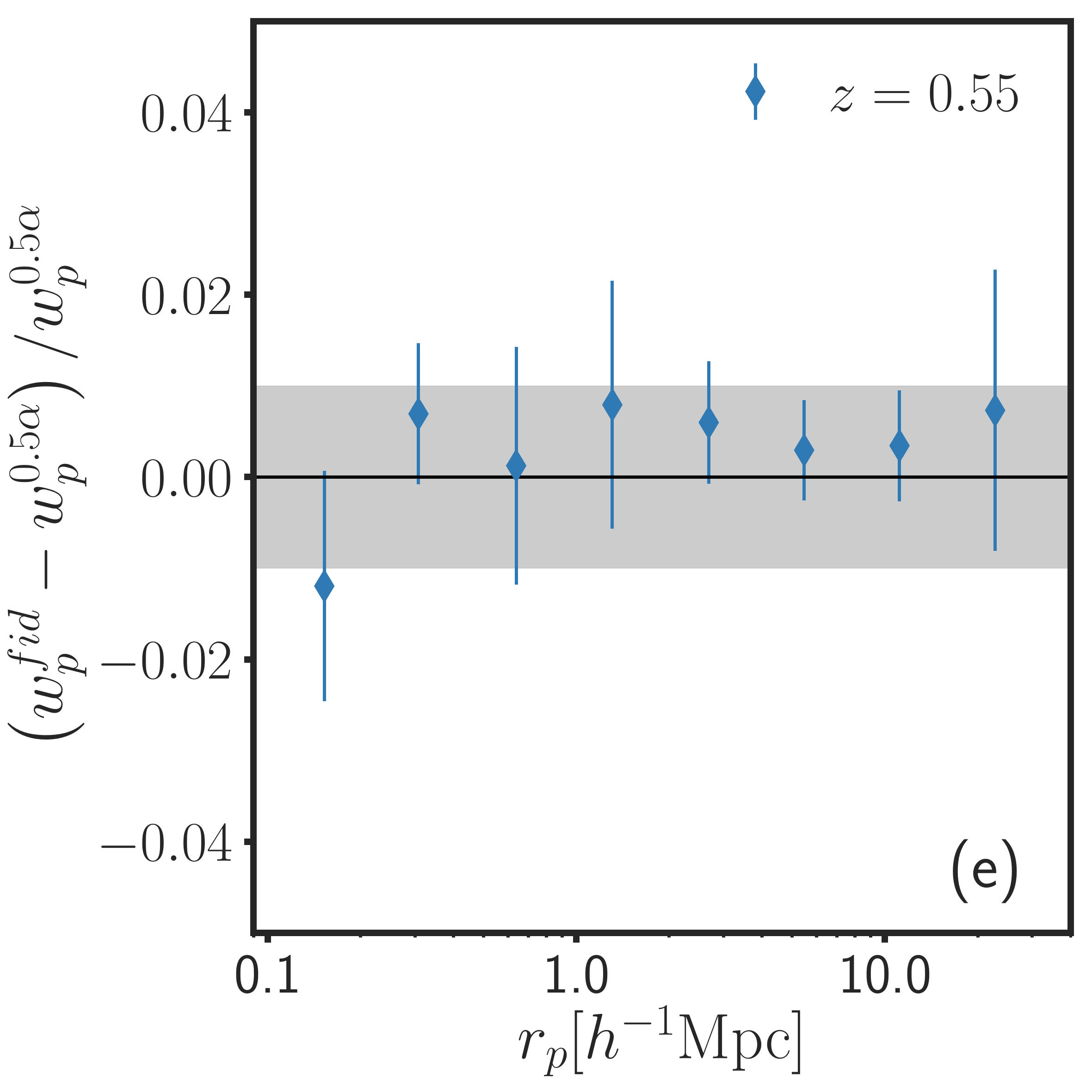} &
  \includegraphics[width=0.3\linewidth]{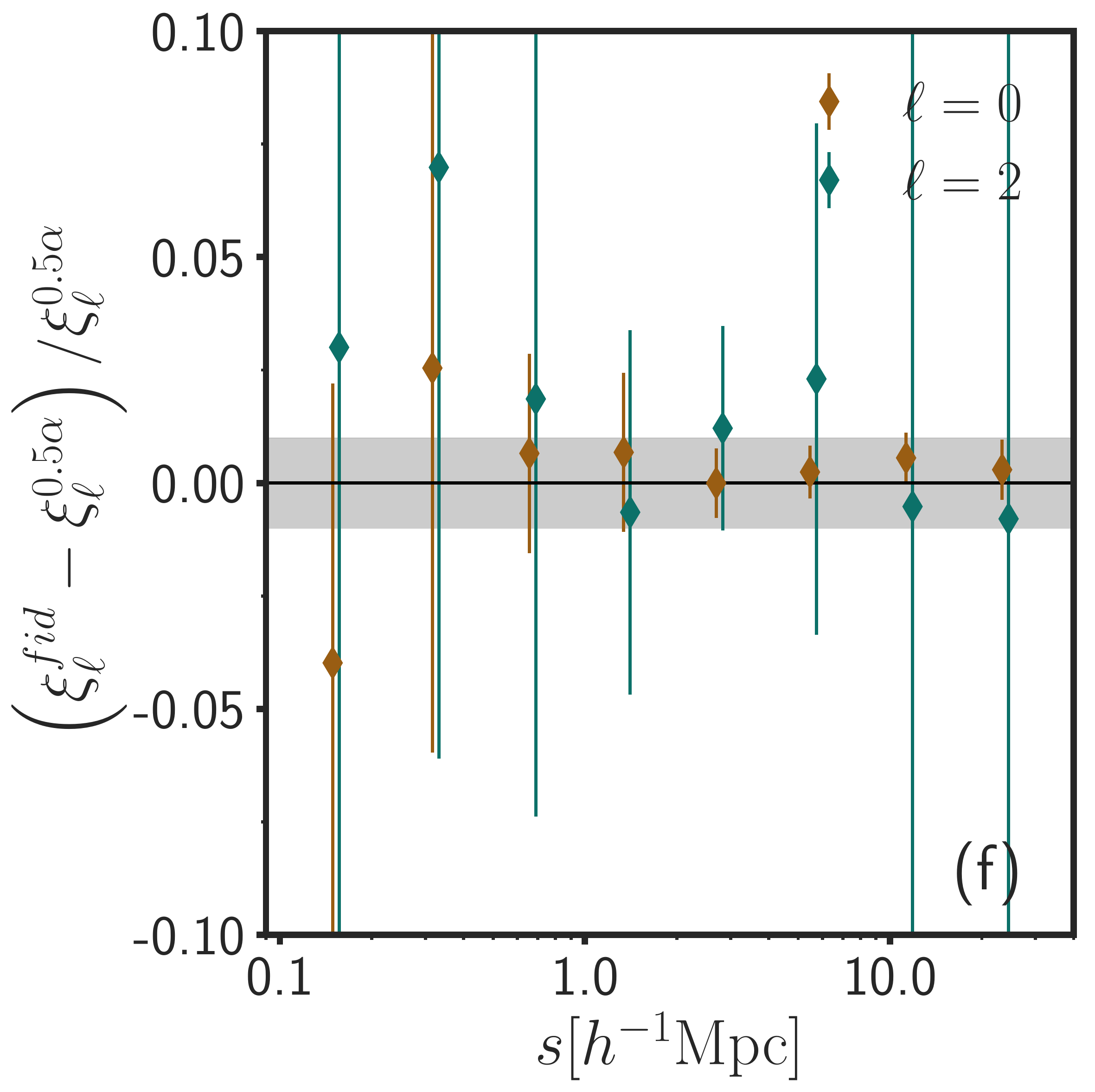}\\   
\end{tabular}
\caption{Convergence tests with respect to force error tolerance. Observables measured from a simulation with our fiducial parameters, \textsc{CT00-Chinchilla} are compared to \textsc{CT30}, a simulation with half the force error tolerance: $\alpha=0.002$. Subfigures are the same as in \autoref{fig:starttime}.}
\label{fig:forceerrortol}
\end{figure*}

In order to test that our results are converged with respect to this choice, we have run an additional simulation, \textsc{CT30}, with $\alpha=0.001$. We find no significant deviations from convergence as can be seen in \autoref{fig:forceerrortol}.

\subsubsection{Time Stepping}
Another significant choice that must be made in the \textsc{l-gadget2} algorithm is the maximum allowed time step. The time step for the leapfrog integrator that \textsc{l-gadget2} uses is determined by $\Delta \textrm{ln}(a) = \textrm{min}\left [ \Delta \textrm{ln}(a)_{\textrm{max}}, \sqrt{2\eta \epsilon /|a|}\right ]$ where $\eta$ is the free parameter determining integration error tolerance. Typically $\Delta\textrm{ln}(a)_{\textrm{max}}$ sets the time step at early times when densities are low, and the $\sqrt{2\eta \epsilon /|a|}$ criterion sets the time step in collapsed regions at late times and thus is important in dictating the convergence of halo density profiles \citep{Power2003}. 

\begin{figure*}[htbp!]
\begin{tabular}{ccc}
\centering
  \includegraphics[width=0.3\linewidth]{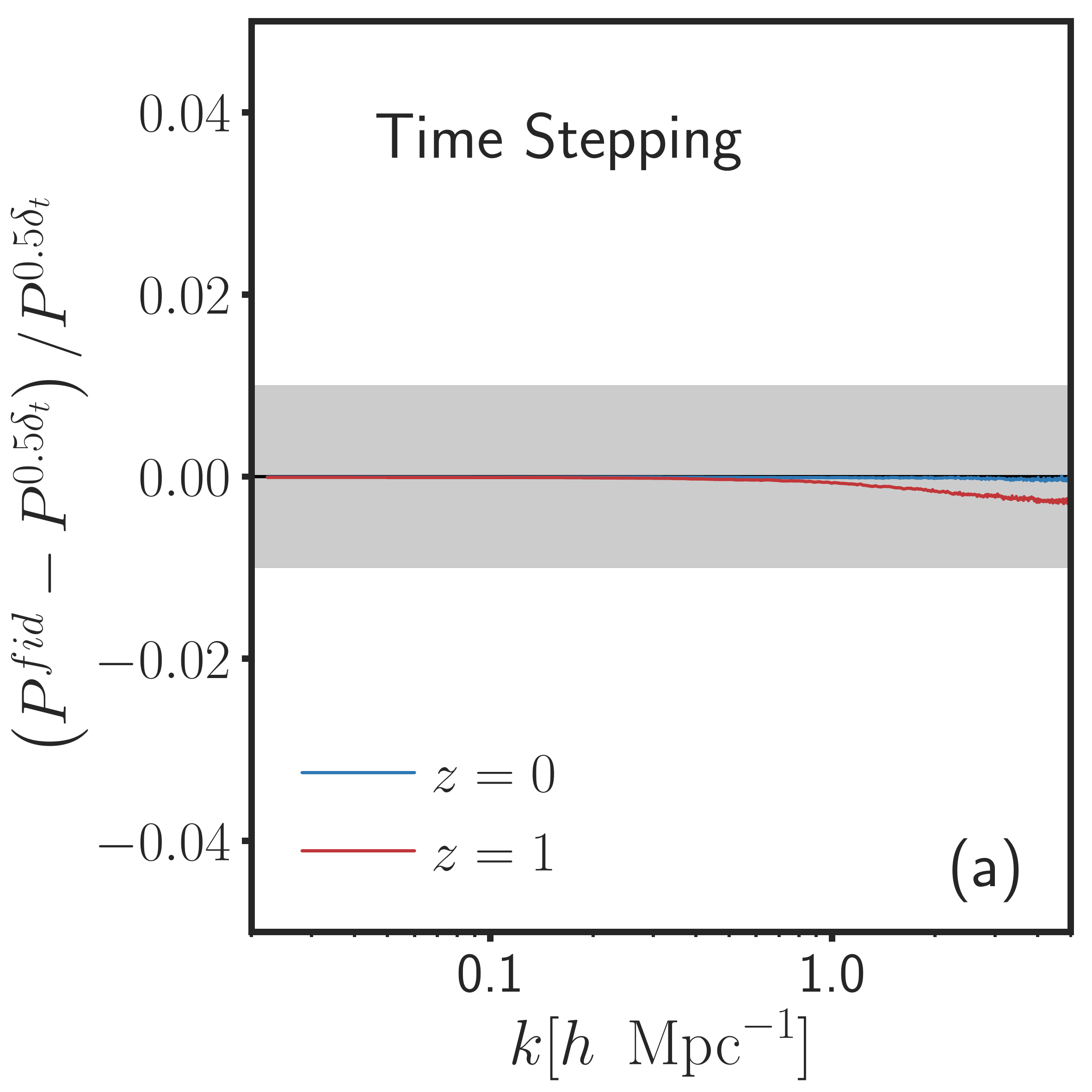}  &
 \includegraphics[width=0.3\linewidth]{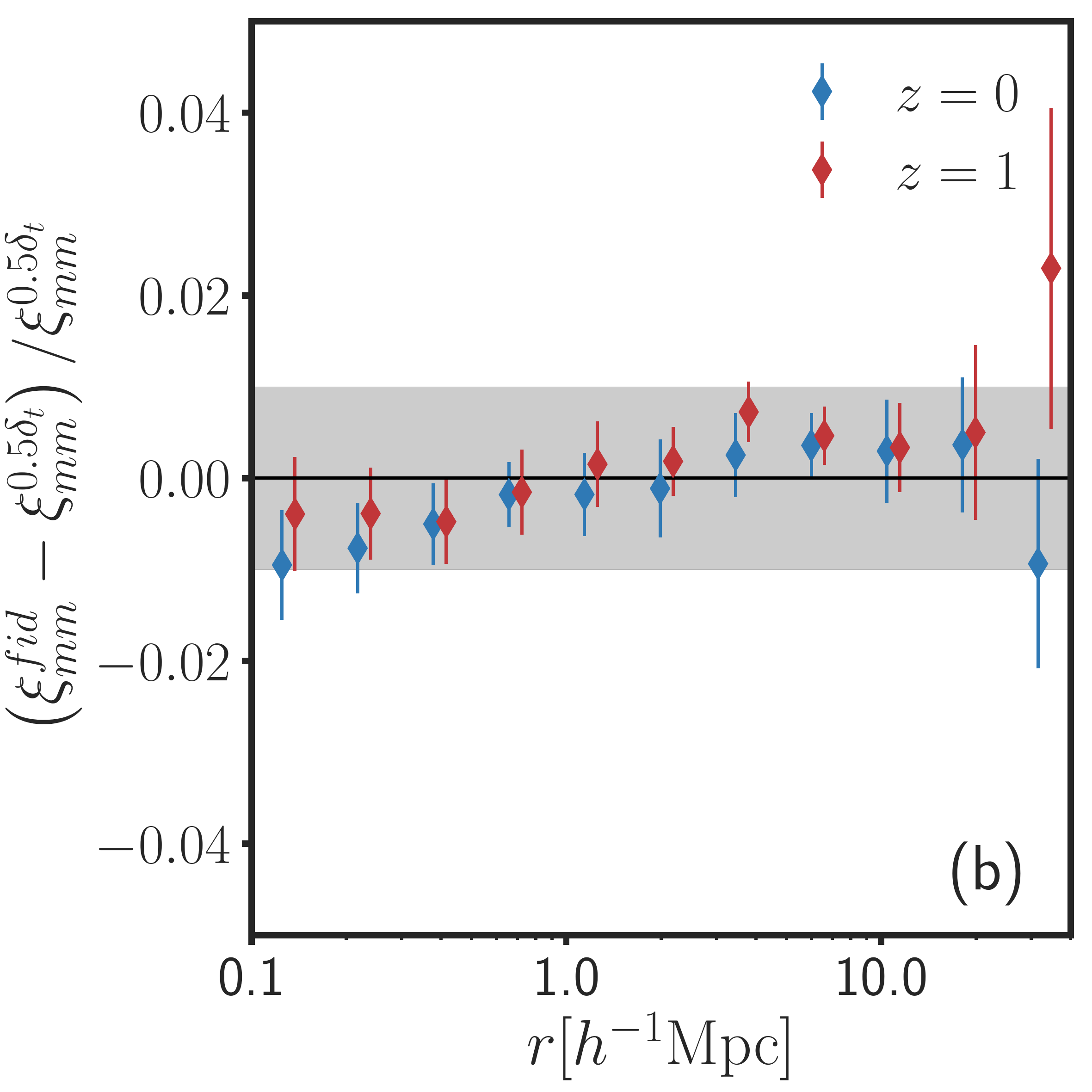} &  
  \includegraphics[width=0.3\linewidth]{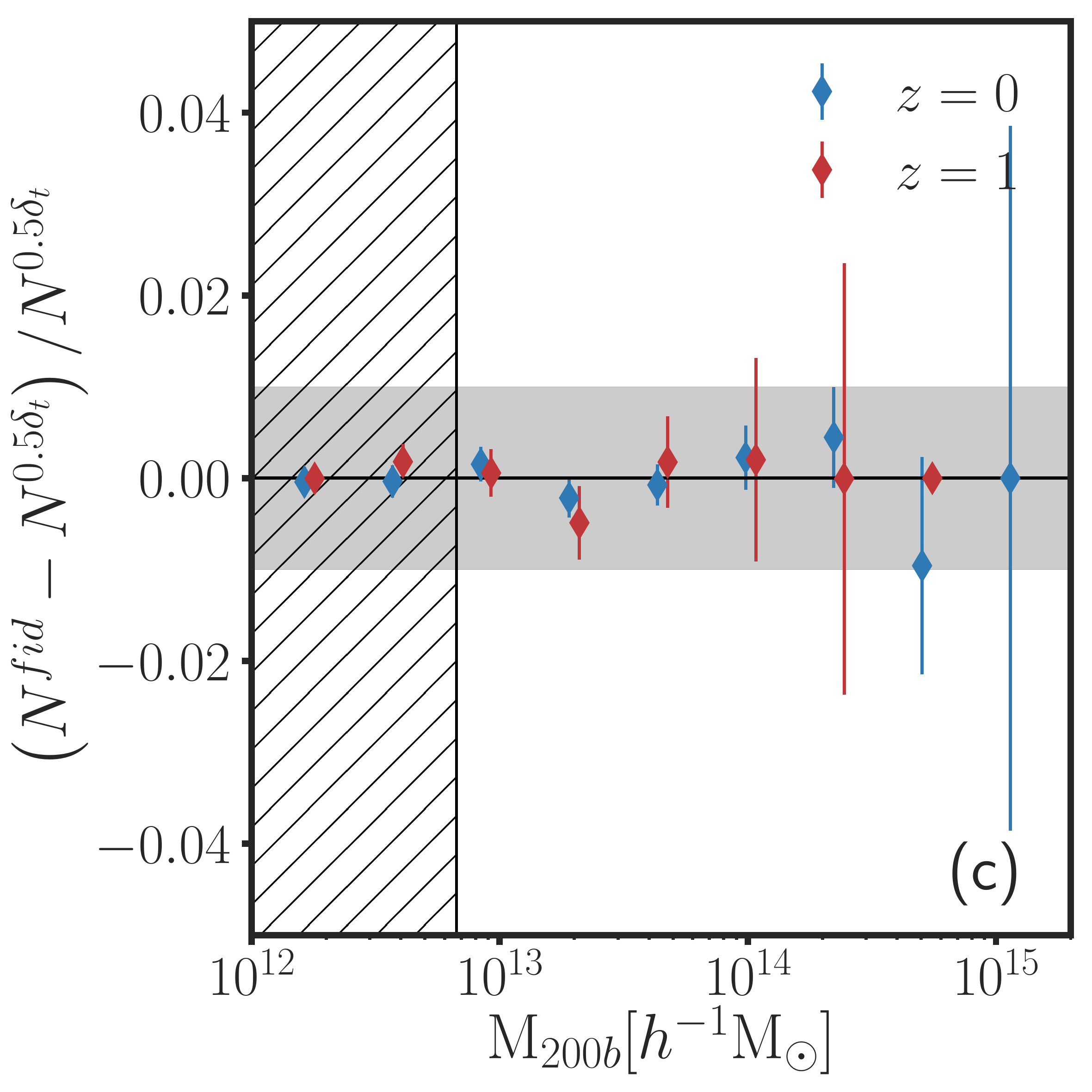} \\
  \includegraphics[width=0.3\linewidth]{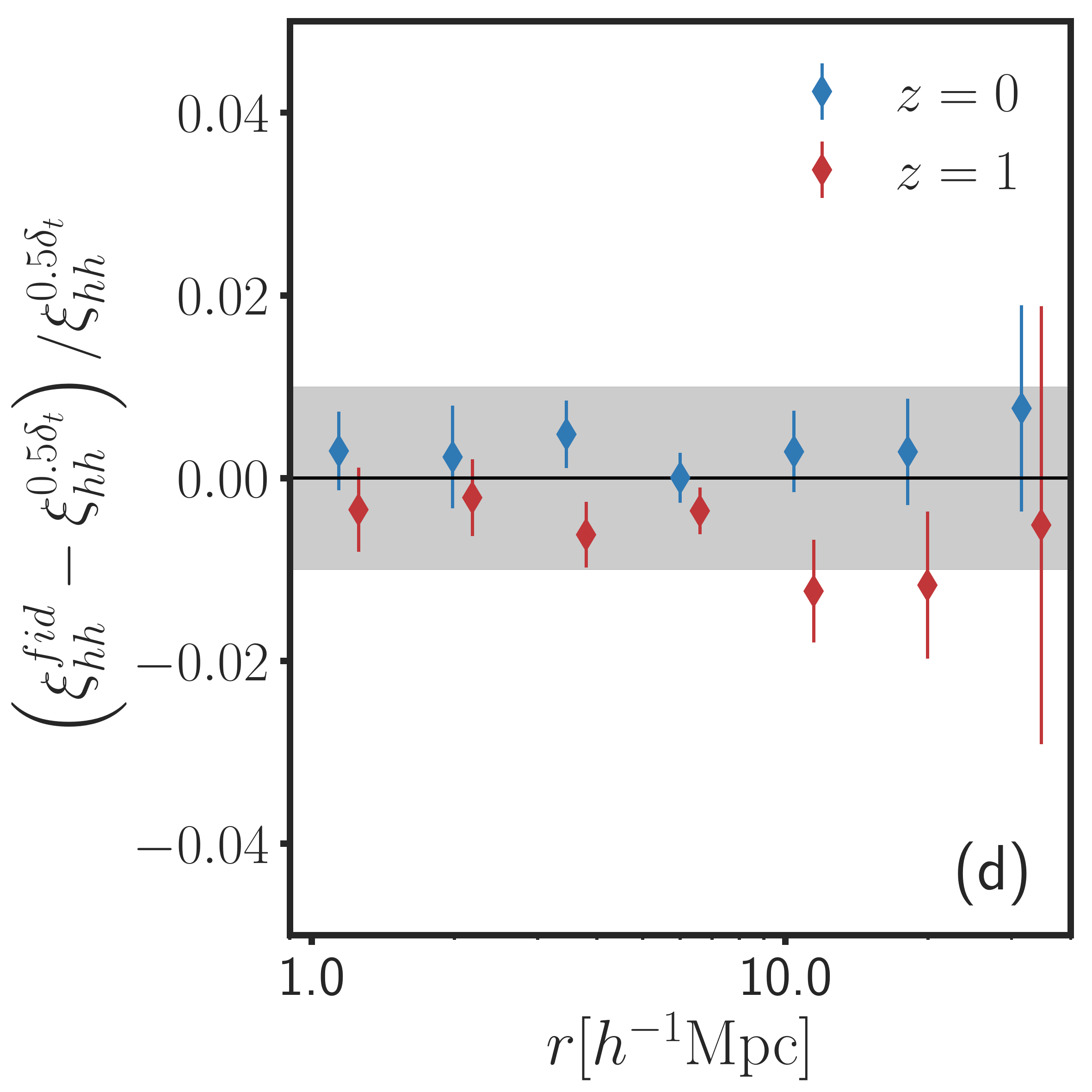}  & 
  \includegraphics[width=0.3\linewidth]{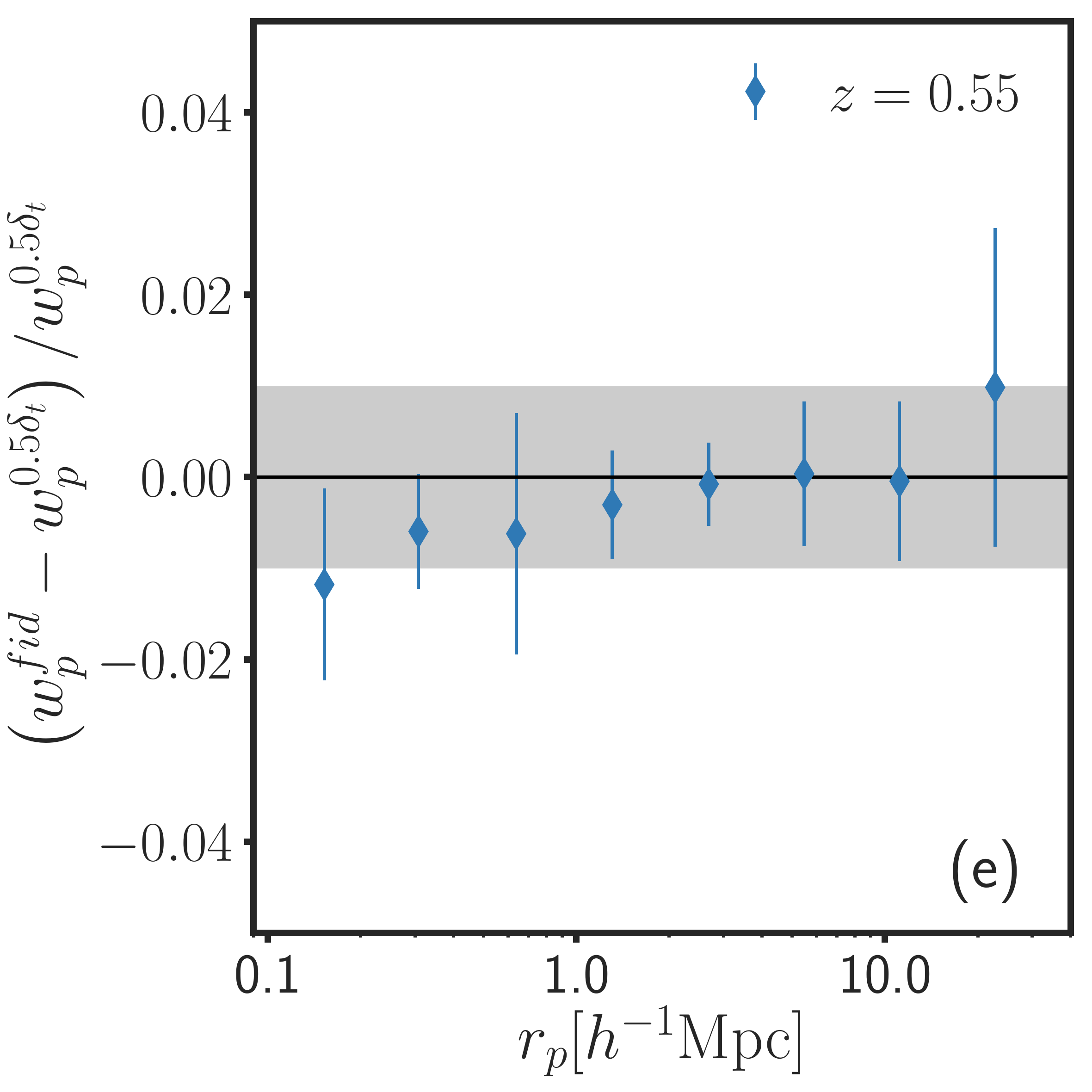} &
  \includegraphics[width=0.3\linewidth]{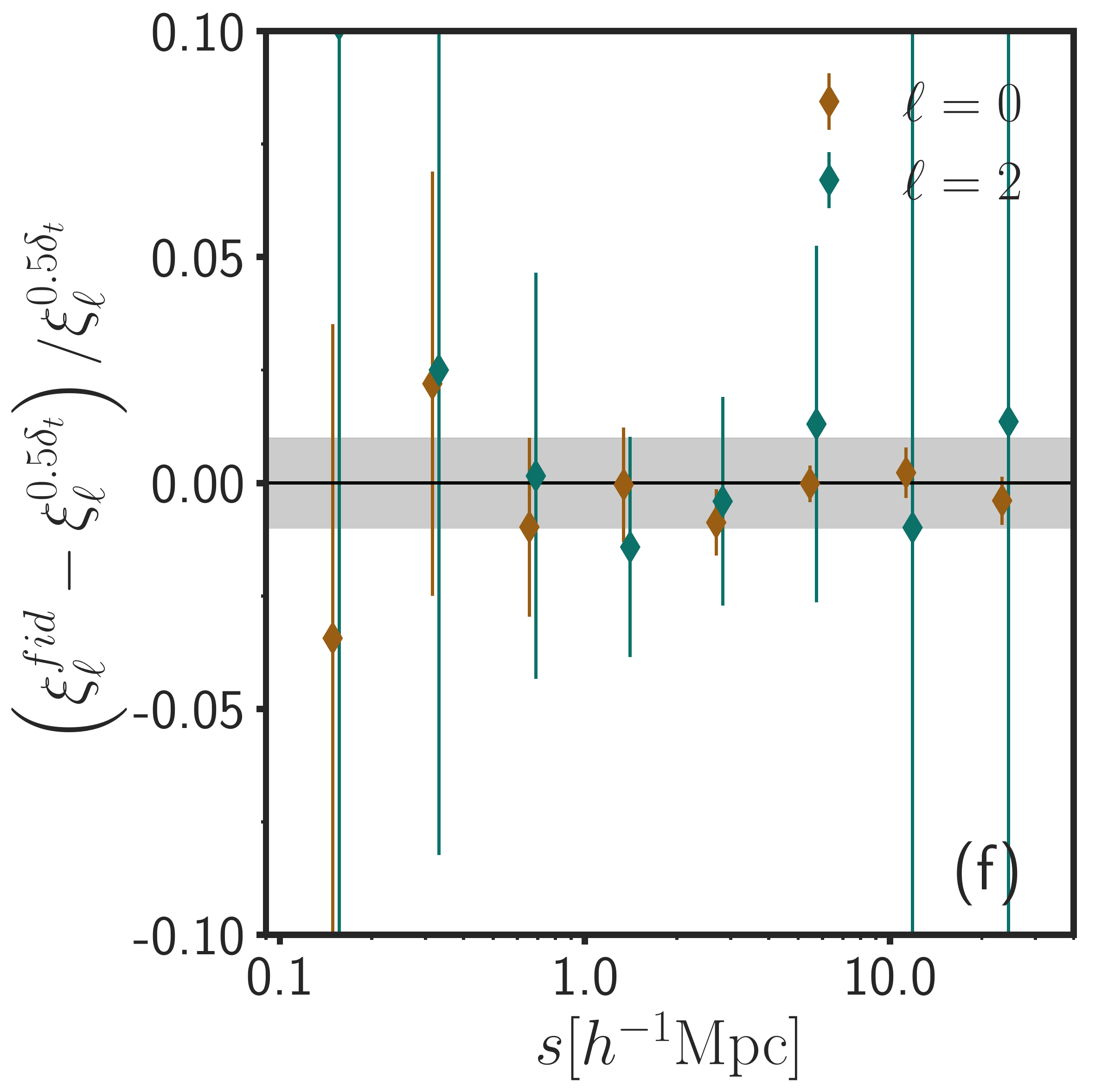} \\  
\end{tabular}
\caption{Convergence tests with respect to maximum time step. Observables measured from a simulation with our fiducial parameters, \textsc{CT00-Chinchilla} are compared to \textsc{CT40}, a simulation with half the maximum time step: $\Delta \ln(a)_{\mathrm{max}}=0.0125$. Subfigures are the same as in \autoref{fig:starttime}.}
\label{fig:timestepping}
\end{figure*}

We have run additional simulations in order to check convergence with respect to time-stepping criteria. In \textsc{CT40}, $\Delta\textrm{ln}(a)_{\textrm{max}}=0.0125$ and in \textsc{CT50}, $\eta=0.0125$, half of their respective values for our fiducial simulation, \textsc{CT00}. Comparisons of the same measurements detailed in \autoref{subsec:measurements} between \textsc{CT00-Chinchilla} and \textsc{CT40} are shown in \autoref{fig:timestepping}. No significant deviations are found. The same comparisons were made between \textsc{CT00-Chinchilla} and \textsc{CT50} and were found to be nearly identical, and so we have not included them for conciseness. 

\subsubsection{Particle Loading}
\label{subsubsec:particleloading}
\begin{figure*}[htbp!]
\begin{tabular}{ccc}
  \includegraphics[width=0.3\linewidth]{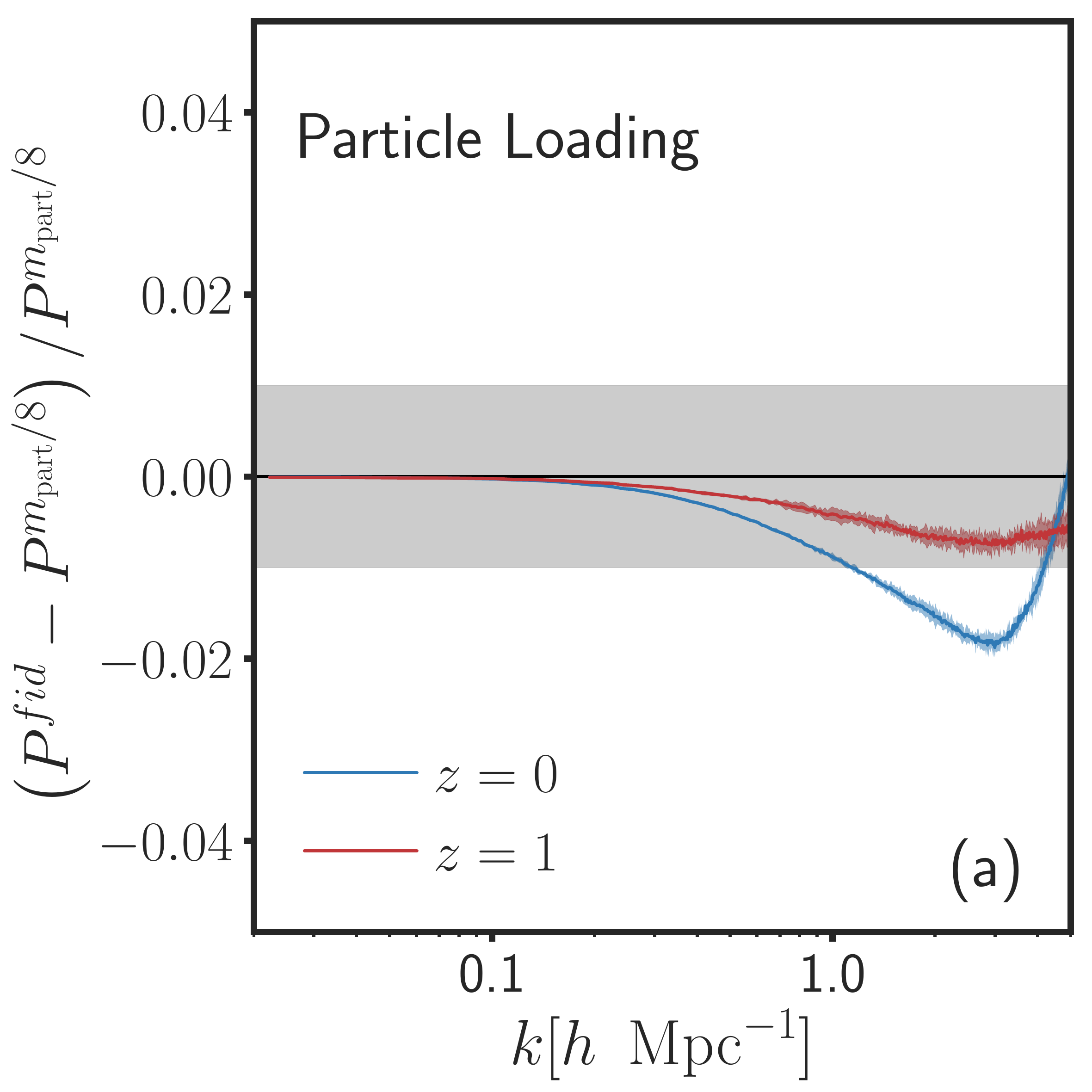} &
  \includegraphics[width=0.3\linewidth]{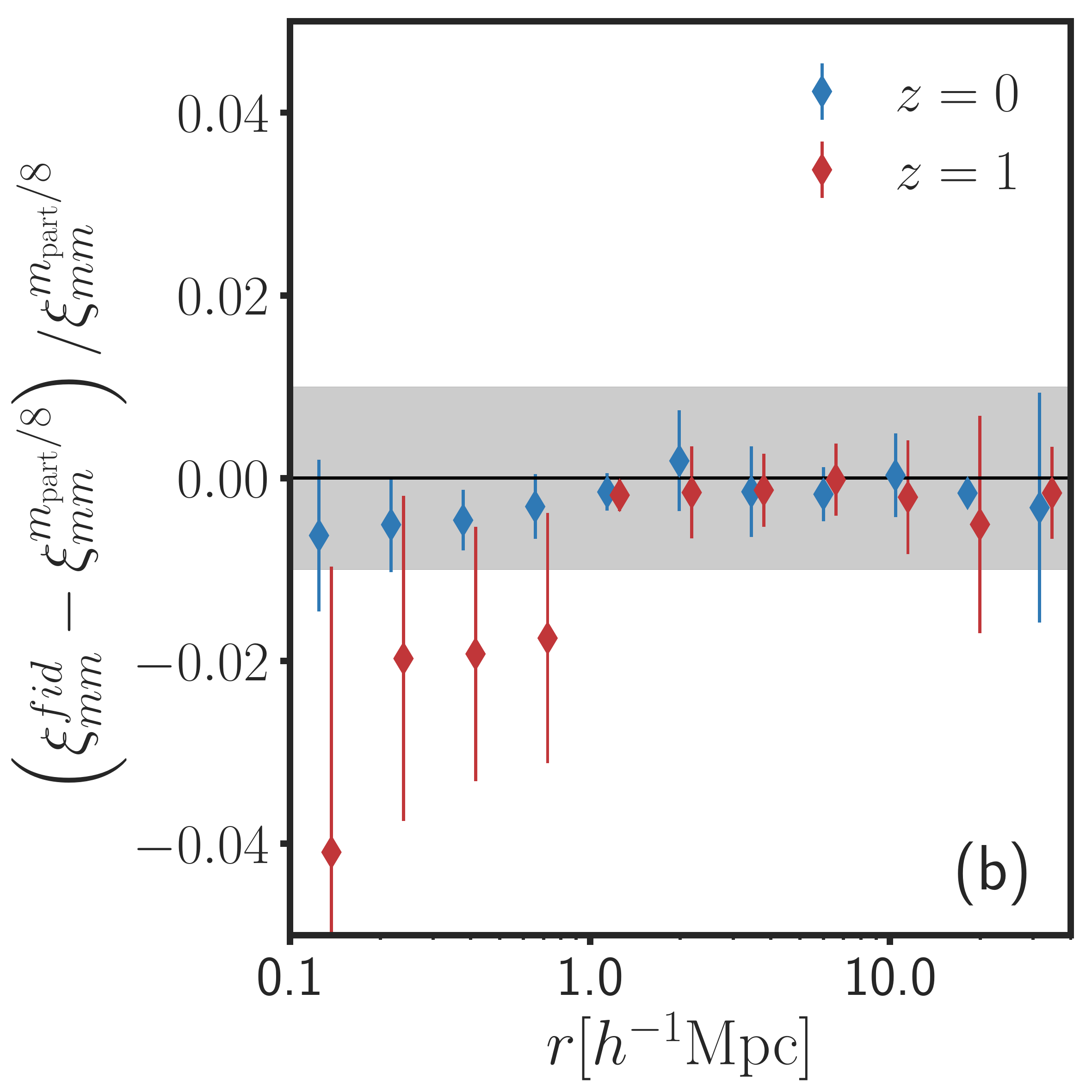}  &  
  \includegraphics[width=0.3\linewidth]{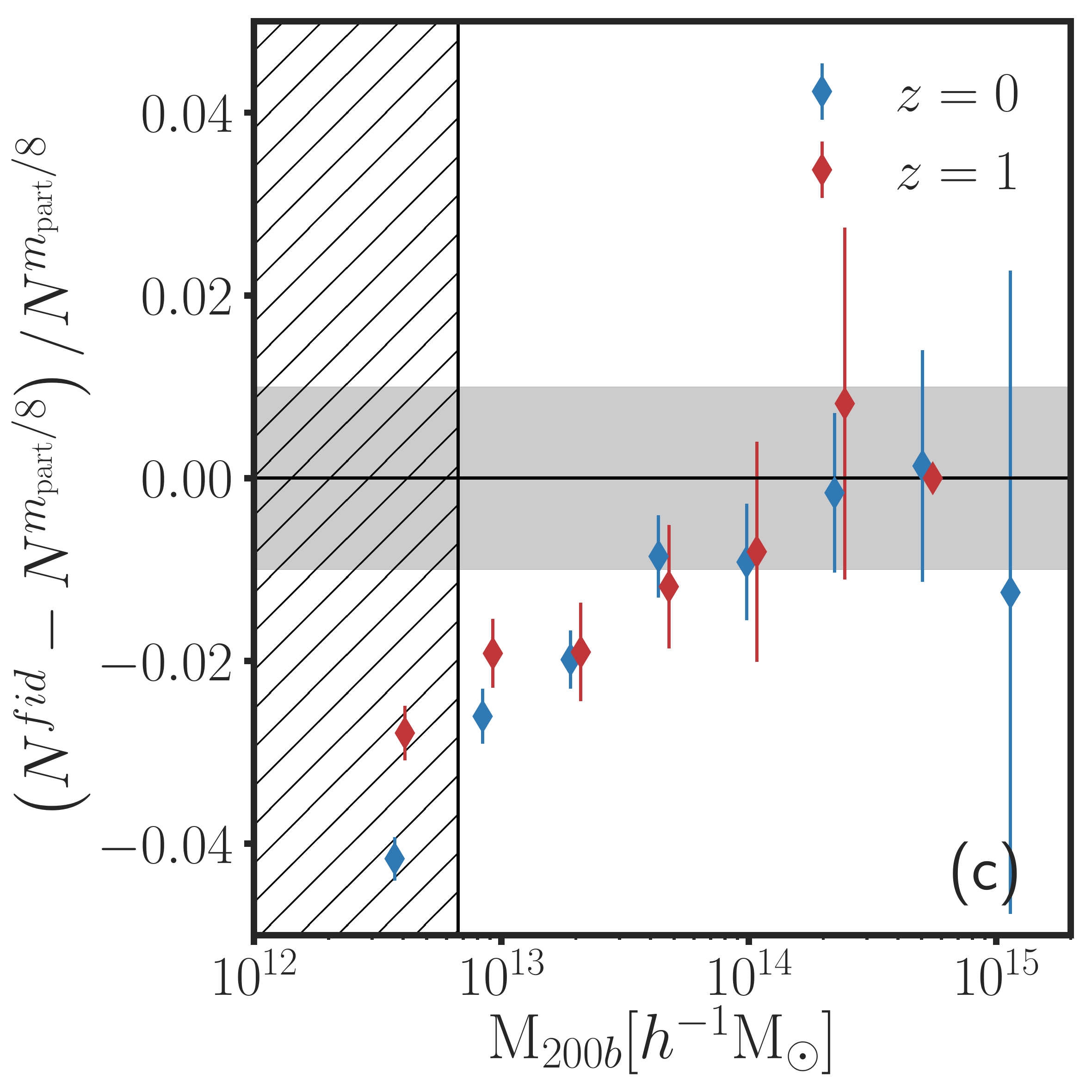} \\ 
  \includegraphics[width=0.3\linewidth]{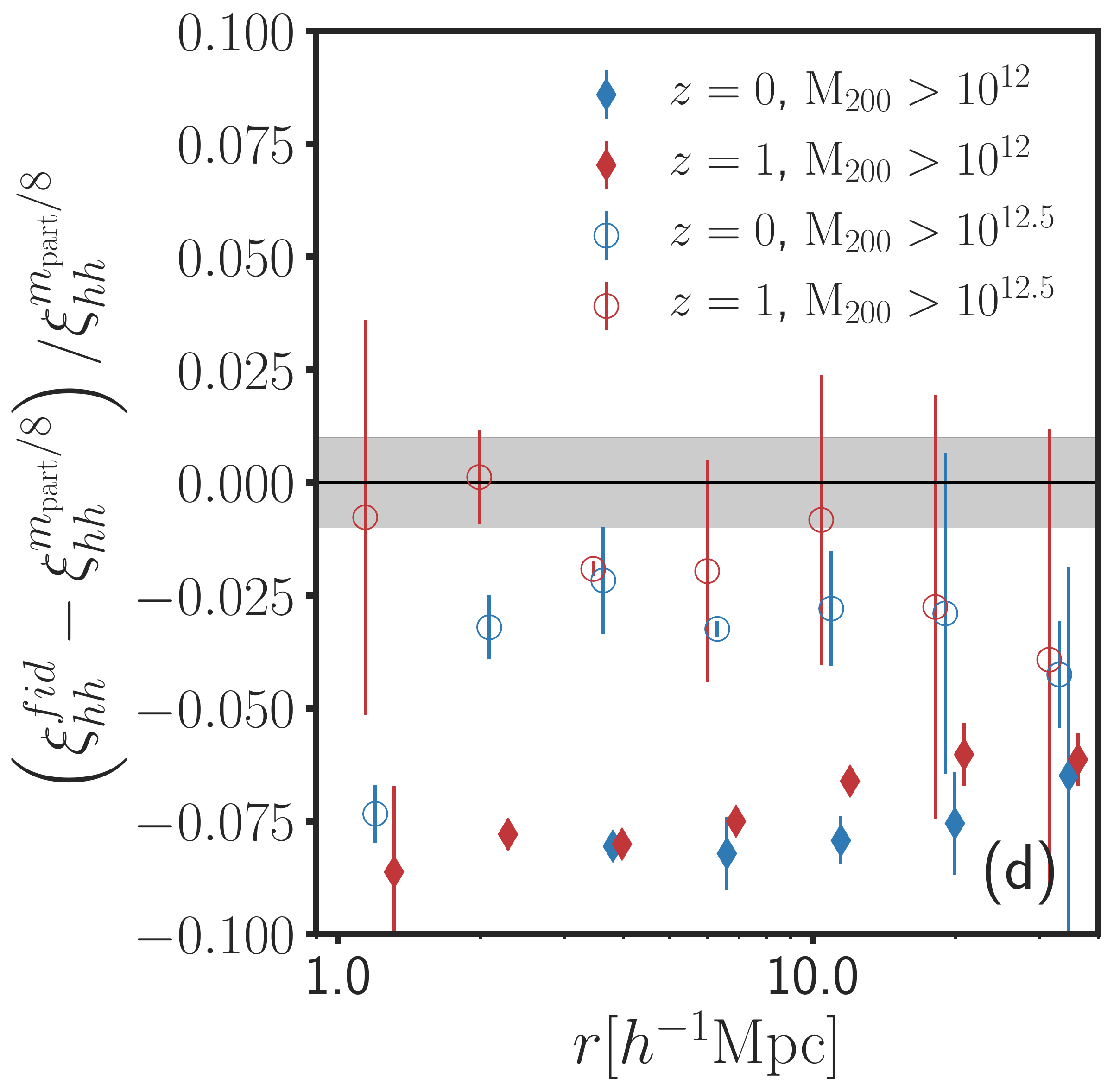}  &
  \includegraphics[width=0.3\linewidth]{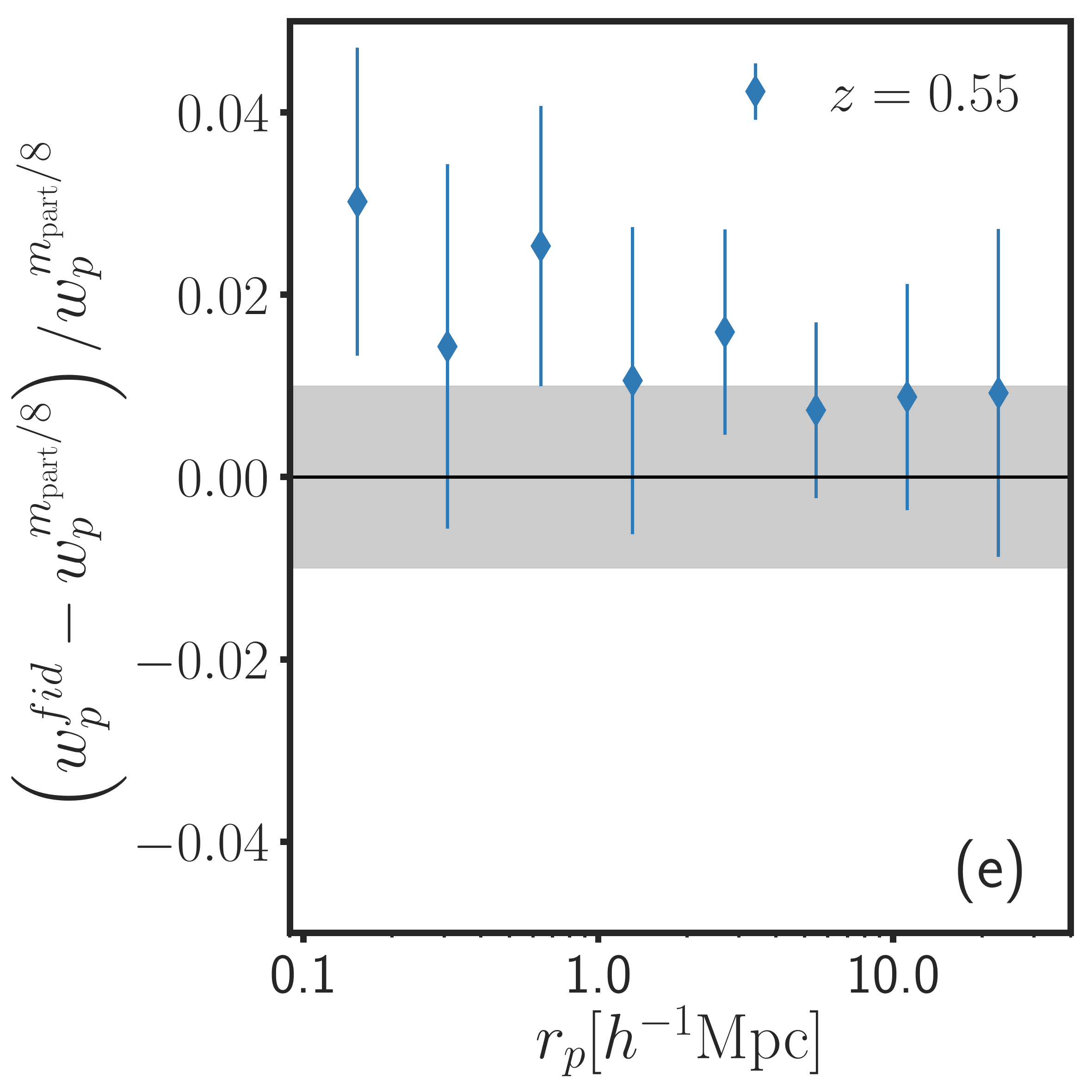} &
  \includegraphics[width=0.3\linewidth]{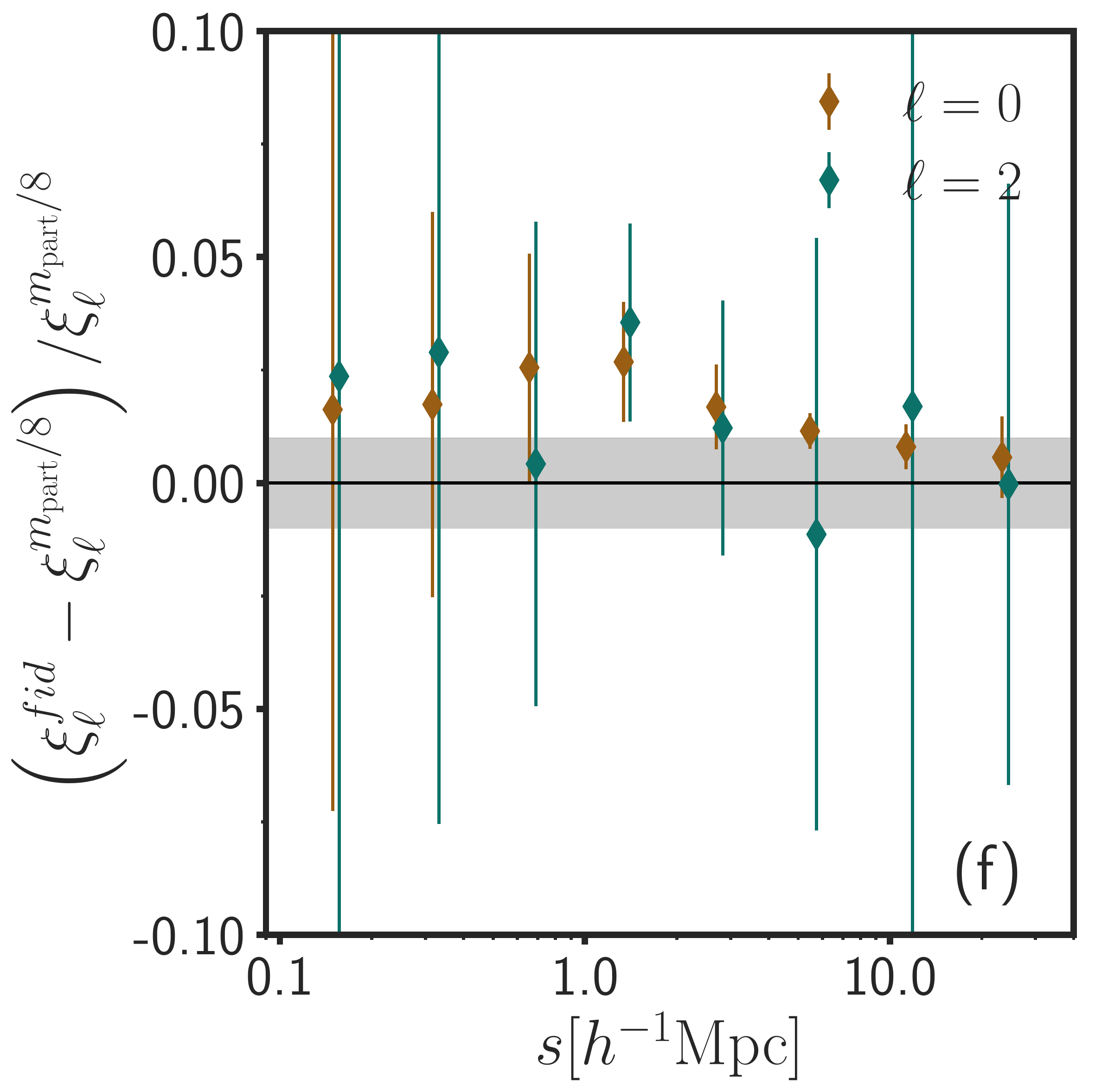} \\   
\end{tabular}
\caption{Convergence tests with respect to mass resolution. Subfigures are the same as in \Cref{fig:starttime}}.
\label{fig:massresolution}
\end{figure*}

The \nbody\ algorithm solves for the evolution of a discretization of the phase-space distribution function of dark matter. Since this phase-space distribution is fundamentally continuous, at least on macroscopic scales, an important parameter governing the accuracy of the algorithm is the number of particles used to sample this distribution function. For the following tests we have run a set of simulations, \textsc{CT60,...,CT63} in three different cosmologies, \textsc{Chinchilla}, \textsc{T00}, and \textsc{T04}, where we have doubled the number of points with which we sample each spatial dimension, increasing the mass resolution by a factor of 8 from our fiducial settings. Results for the comparison of these simulations with our fiducial set in the Chinchilla cosmology can be found in \autoref{fig:massresolution}. 

\begin{figure}[htbp!]
  \includegraphics[width=\columnwidth]{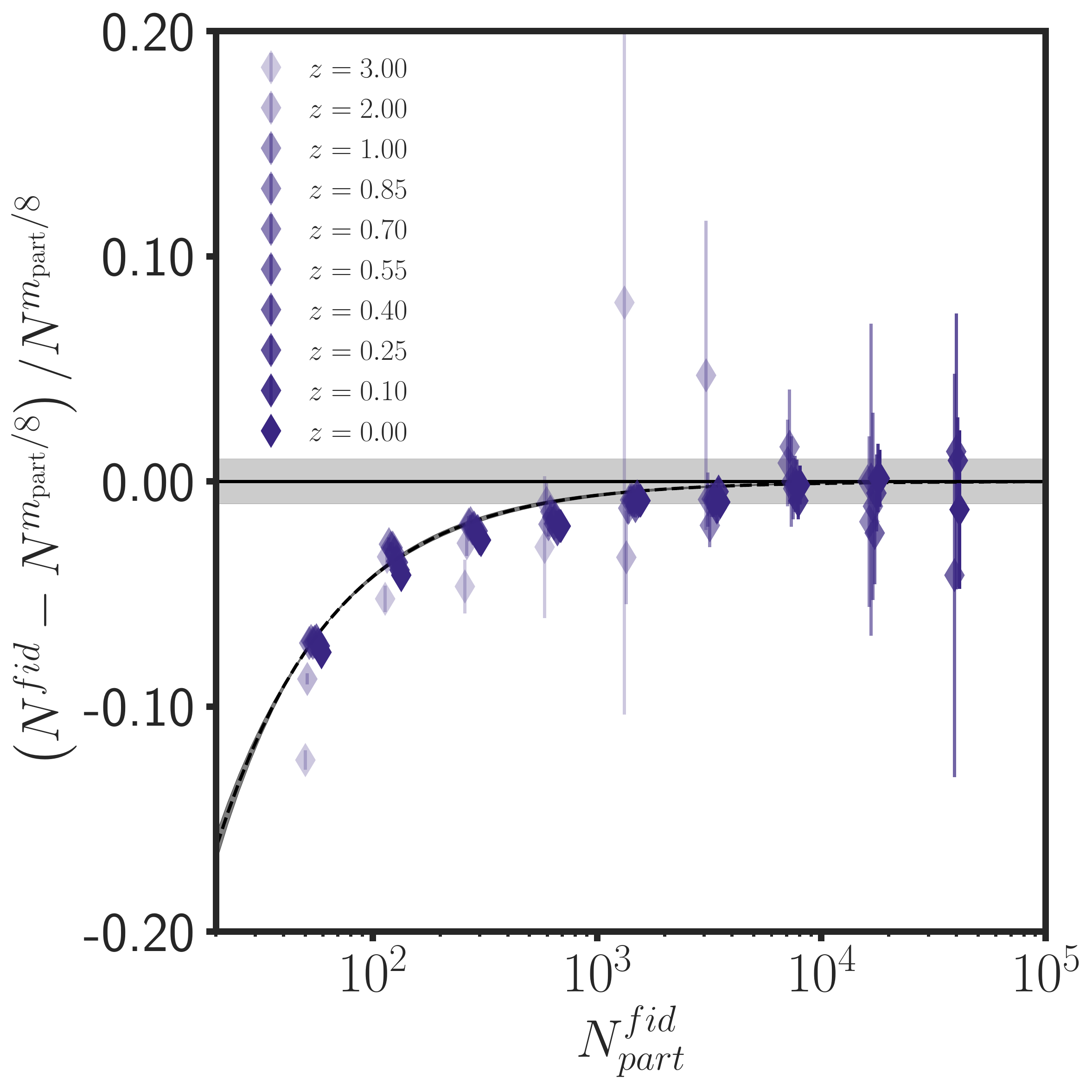} \\
\caption{Deviations of the mass functions measured in simulations using our fiducial parameters from simulations with higher mass resolution as a function of redshift. The line is a fit to all of these points in addition to the points for the other two cosmologies that are not shown in this figure.}
\label{fig:massresolutionfit}
\end{figure}

The mass function is converged to within $1\%$ for halos that are resolved with 500 particles or more. For masses below this, we observe varying degrees of deviation from convergence which depend to good approximation on just the number of particles that the halo is resolved with. This can be seen in \autoref{fig:massresolutionfit}, which demonstates that bins in particle number show similar behavior for all redshifts except for very low particle numbers at high redshift. Only one cosmology is plotted, but a similar trend holds in the other two cosmologies, despite the three cosmologies spanning a large range in $\sigma_{8}$ and $\Omega_{m}$. We have fit the following function to the average of these residuals over redshift in order to characterize and correct for them in other works:

\begin{align}
\label{eq:massresfit}
\frac{N(M^{fid}_\text{200b}) - N(M^{m_{\mathrm{part}}/8}_\text{200b})}{N(M^{m_{\mathrm{part}}/8}_\text{200b})} &= -\textrm{exp}\frac{-(\log_{10}N_{\text{part}}-\log_{10}N_{0})}{\sigma_{\log_{10}N}}
\end{align}
where $N_{\mathrm{part}}$ is the number of particles corresponding to $M^{fid}_\text{200b}$, the halo mass measured in our fiducial simulations. We find $\log_{10}N_{0}=0.25\pm0.13$ and $\sigma_{\log_{10}N}=0.557\pm 0.046$. 

\begin{figure}[htbp!]
  \includegraphics[width=\columnwidth]{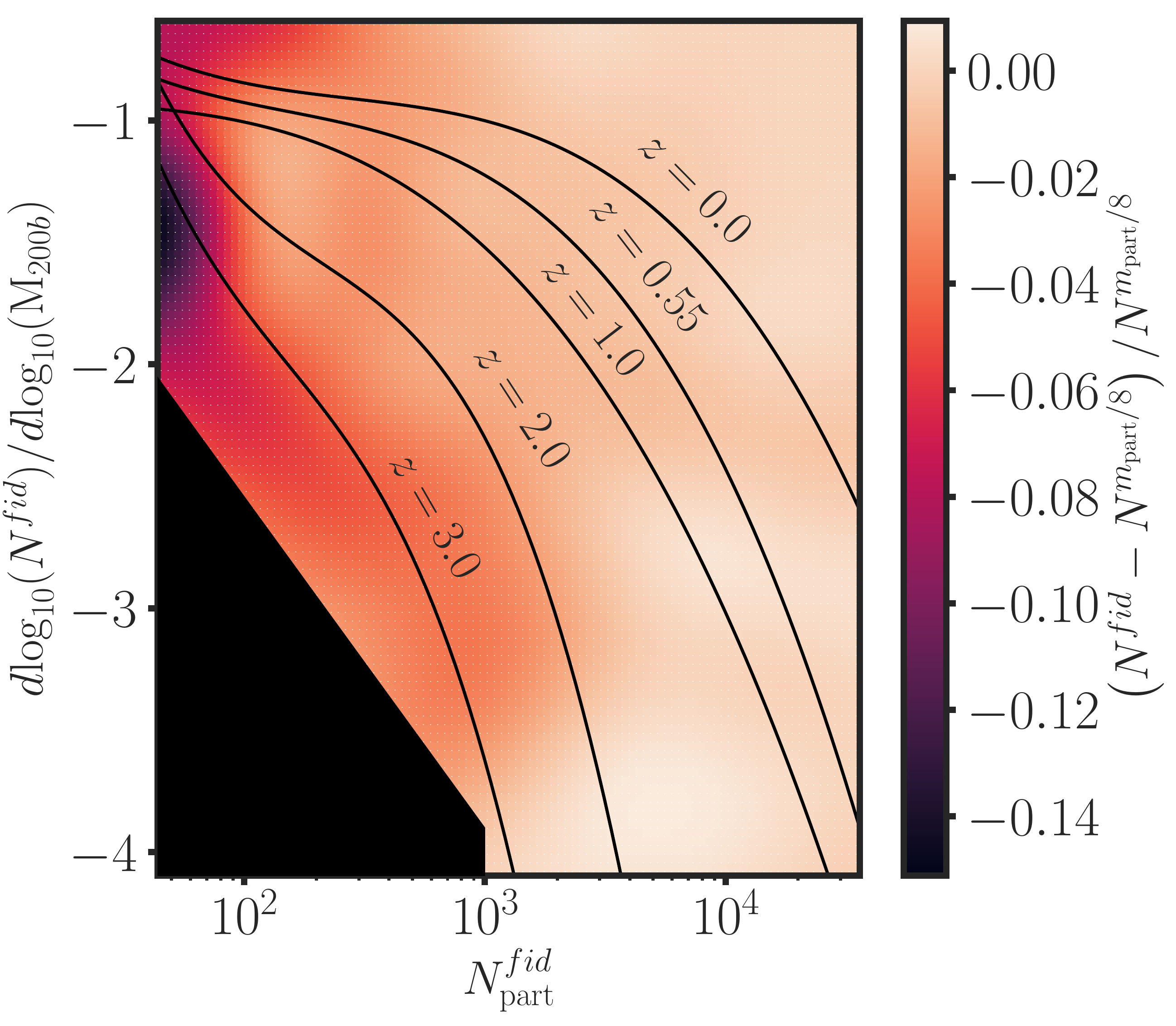} \\
\caption{Deviations of the mass functions measured in simulations using our fiducial parameters from simulations with higher mass resolution as a function of particle number and logarithmic slope of the mass function. A clear trend can be seen with logarithmic slope of the mass function for $N_{\mathrm{part}}<1000$. Black lines show the logarithmic slopes of mass functions measured in the \textsc{CT00-T00} simulation for different redshifts. The solid black region is beyond where we have any data and so we exclude it from this plot.}
\label{fig:logslope}
\end{figure}

To higher order, the deviations from convergence appear to be dependent on the local logarithmic slope of the mass function, $\Gamma = \frac{d \log_{10} N^{fid}}{d\log_{10}M_\text{200b}}$, with the worst deviations occurring at low particle number and very steep slopes. This can be seen in \autoref{fig:logslope}. Here, we have measured the deviations of our fiducial simulations from convergence as a function of particle number and $\Gamma$, where $\Gamma$ is determined by fitting a quartic spline to $N(M_\text{200b})$ in the \textsc{CT0} simulations at all redshifts and taking its logarithmic derivative. We have also interpolated these measurements in order to make the trends more obvious. Above about 1000 particles, the deviations from convergence of the mass function are less than $1\%$ for all slopes. Below this particle number, there is a trend in error with $\Gamma$, leading to the larger errors seen at high redshift in \autoref{fig:massresolutionfit}. For these reasons, we caution against using the correction as determined above for halos with particle numbers less than 1000 when $\Gamma<-2$. 

The deficit of halos that we find in our fiducial simulations compared to the \textsc{CT6} simulations cannot be explained by increased Poisson random noise in the mass estimates, as this would lead to an over-abundance of halos at a given mass due to the negative slope of the mass function in a manner analogous to Eddington bias. Instead, the observed deficit suggests that a bias is being introduced in the density field, which is clear from the deviations observed in $P(k)$, such that low mass halos are less likely to form in lower resolution simulations. 

These errors also propagate into other observables involving halo mass. For instance, $\xi_{hh}$ deviates from convergence by $7.5\%$ when using all halos with $M_\text{200b}>10^{12} ~\hmsun$, but quickly converges as a function of mass as can be seen by the fact that halos with $M_\text{200b}>10^{12.5} ~\hmsun$ only deviate by $3\%$ from the simulations with higher particle loading. Mass cuts above this have noisy $\xi_{hh}$ measurements and so we cannot make precise statements about their convergence. The galaxy correlation functions are less sensitive to mass resolution at the low mass end because our HODs are tuned to match the massive BOSS massive galaxy sample. This can be seen in  \autoref{fig:massresolution}e and \autoref{fig:massresolution}f, where $w_{p}$ is converged at the $1-2\%$ level, with the redshift space measurements performing only slightly worse. $\xi_{mm}$ is converged, while $P(k)$ deviates from convergence for $z=0$ above $k\sim 1.5\/\invhmpc$, with a maximal deviation of about $2\%$ at $k\sim 3\/\invhmpc$. The deviations from convergence for $P(k)$ are consistent with those found in \citet{Schneider2016} who find $\sim 1\%$ deviations from convergence for $P(k)$ at $k\sim 1\invhmpc$ for a $L_\text{box}=512$, $N_\text{part}=512^3$ simulation.

\subsubsection{Finite Box Effects}
\begin{figure}[htbp!]
  \includegraphics[width=\columnwidth]{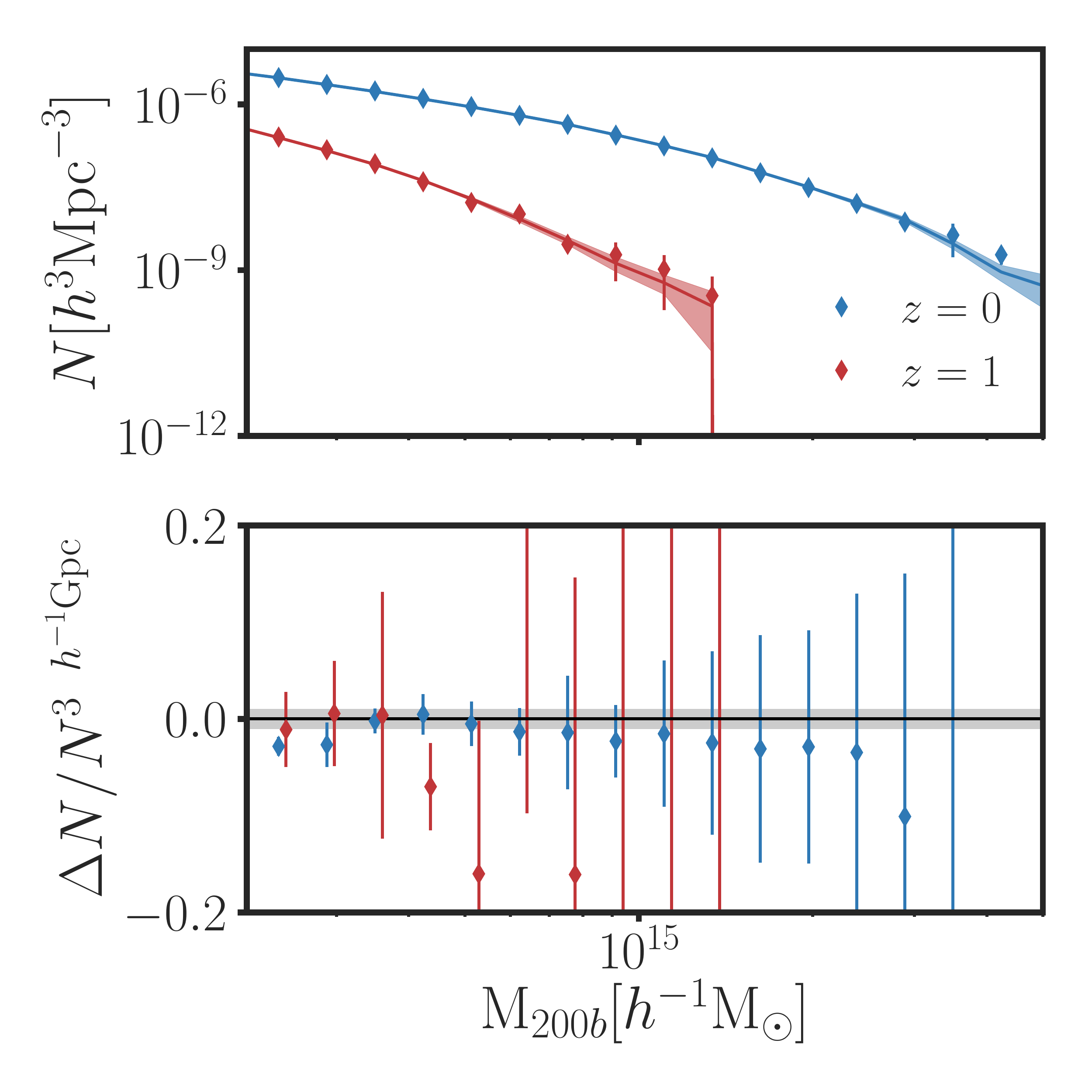} \\
\caption{Comparison of the average mass function over the five realizations of \textsc{T04} with that measured in \textsc{CT7-T04}. The top panel shows the mass functions, where the points are measured from \textsc{T04} and the lines are measurements from \textsc{CT72-T04}. The bottom panel shows the fractional difference of the mass functions between these simulations. These measurements are consistent with no finite box effects.} 
\label{fig:finitebox}
\end{figure}

It is currently beyond the realm of possibility to simulate the entire observable universe at high enough resolution to be useful. Instead, the common practice in cosmological simulations is to assume periodic boundary conditions with a fundamental mode which is much larger than the scales of interest for the problem at hand. One effect of doing this is that modes larger than the fundamental mode of the box are not included in the growth of structure. Because gravitational collapse is a non-linear process, the growth of small-scale structure couples to large-scale growth, and thus missing large variance can cause inaccuracies and alter sample variance at smaller scales. Additionally, because our simulations are periodic, only discrete modes, $\vec{k}=\frac{2\pi(i,j,k)}{L_{box}}$ where $i,j,k\in \mathbb{Z}$ are included in the initial conditions. In order to test the effects of these approximations we have run a set of much larger, lower resolution simulations, \textsc{CT70,...,CT76} at the same cosmologies as our test simulations, where we have $(5\,\ \hgpc)^3$ for each cosmology to compare with. The results of the comparison for the \textsc{T04} cosmology are shown in  \autoref{fig:finitebox}; the other cosmologies show nearly identical results.

Because the \textsc{CT7} simulations have worse mass resolution than our test simulations, the analysis in \autoref{subsubsec:particleloading} indicates that there should be residual effects in this comparison due to mass resolution. In order to mitigate the differences arising from mass resolution, we have applied the correction in \autoref{eq:massresfit} to both sets of measurements. We find convergence to within sample variance of the test boxes for all masses at both $z=0$ and $z=1$, although this is significantly larger than the percent level at $z=1$ for all masses shown here. We also compared $\xi_{mm}(r)$ and found no deviations from convergence for $r<100 ~\hmpc$.

\section{Halo Finding}
\label{sec:halofinding}
In this section we discuss the sensitivity of our results to choices made with regards to halo finding. We have defined dark matter halos as spherical structures with overdensities of 200 times the background density. This choice is relatively arbitrary, having its basis in simple spherical collapse models which have been shown to be imprecise compared to modern cosmological standards. As such, we discuss the possible impacts that this definition might have on cosmological results obtained from emulators built upon these simulations. Note that we consider the choice of halo finder and the settings used in that halo finder to be part of the mass definition, and as such we do not consider the effects of different halo finders separately.

In the case of using galaxy clustering to constrain cosmology, there is a large literature on how choice of halo definition can impact and possibly bias inferred cosmology. Much of this literature has focused on the effect of secondary parameters on the clustering signals of halos at fixed mass \citep[e.g.][]{Wechsler2006,Gao2005,Mao2018,Chue2018}. This effect propagates differently into galaxy clustering depending on which proxy is then used to assign galaxies to halos \citep{Reddick2013, Lehmann2017}, and can lead to biases in inferred HOD parameters when neglected \citep{Zentner2014}. Whether these effects lead to biases in inferred cosmology when using HODs, and whether these biases can be mitigated through extensions to the HOD model are still open questions.

In the case of the halo mass function, the situation is equally complicated. We do not directly measure the halo mass function, unlike the galaxy correlation function, but rather some distribution of observables such as cluster richness or X-ray temperature. In order to constrain cosmology, a mass--observable relation (MOR) must be obtained, and the calibration of this mass--observable relation must also assume a halo mass definition. It is imperative that the definition used when constraining the MOR and the definition used for the halo mass function be the same in order to obtain unbiased cosmological constraints. If the scatter in the MOR is smaller for a particular mass definition, that definition will yield tighter constraints, but a study of the halo mass definition that minimizes scatter in the MOR is beyond the scope of this paper. A more practical reason for our choice of $\Delta=200\text{b}$ is that it makes the typical radii of cluster mass halos, $\sim0.5-2 ~\hmpc$, significantly larger than the force softening lengths used in our simulations. 

\section{Comparison with Other Emulators}
\label{sec:otheremu}
Having internally validated our simulations, we now compare our measurements to those obtained in other works. Unfortunately, the most precise determination of the matter power spectrum available to date, the Mira--Titan universe emulator \citep{Heitmann2016, Lawrence2017}, does not cover the same parameter space as our simulations. In particular, they do not include $N_{\text{eff}}$ in their parameter space, and varying this can lead to deviations in $P(k)$ on the order of $\sim 10\%$, much greater than the precision at which such a comparison would be relevant. Instead, we have compared our simulations to predictions from the widely used \textsc{Halofit} algorithm \citep{Smith2003,Takahashi2012} which does span our parameter space. Our simulations are not large enough in volume for precision emulation of the matter power spectrum, but nevertheless we can compare our measured matter power spectra to the \textsc{Halofit} predictions for our cosmologies as both an external validation of our simulations and as a further consistency check for the \textsc{Halofit} algorithm. 

\begin{figure}[htbp!]
\includegraphics[width=\columnwidth]{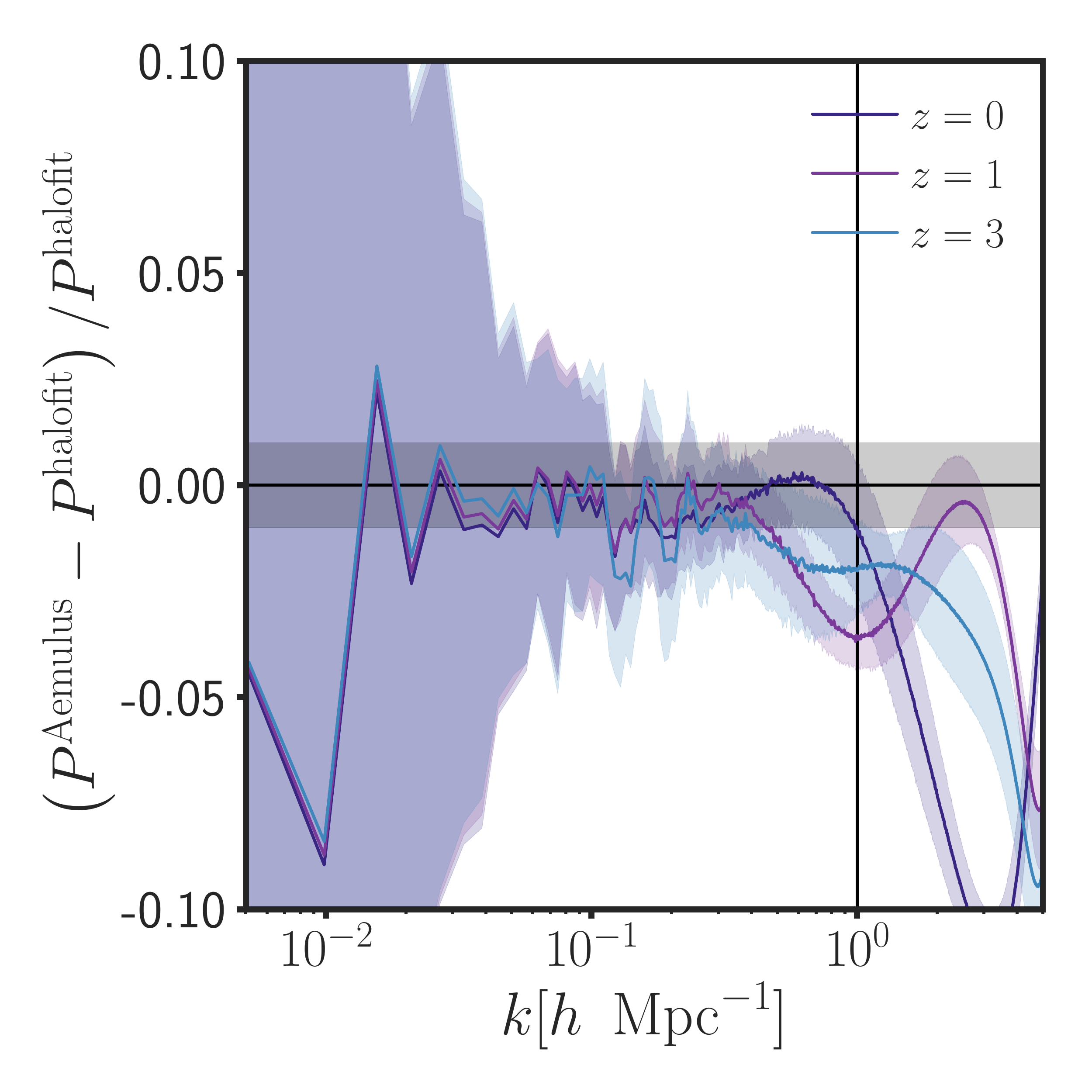} 
\caption{Comparison of the 40 simulation boxes to Takahashi halofit. The vertical black line demarcates the wavenumber that we expect the effects of mass resolution to become important at the $>1\%$ level. Agreement is to within the reported \textsc{Halofit} accuracy.}
\label{fig:halofit_comp}
\end{figure}

The results of this comparison can be found in  \autoref{fig:halofit_comp}. Error bars in this figure correspond to the variance of the deviations of our 40 training simulations from \textsc{Halofit}. We find better than $1\%$ agreement in the mean deviation until $k\sim 0.3~\invhmpc$, but observe maximum errors close to $5\%$, consistent with the \textsc{Halofit} internal error estimation in \citet{Takahashi2012}. For scales smaller than $k=1~\invhmpc$ we find large deviations of up to $12\%$ which are likely due to a combination of inaccuracies in \textsc{Halofit} and resolution effects in our simulations. The maximum errors that we observe for $0.1<k<1$ are slightly smaller than those reported in \citet{heitmann2014}, but this may be attributable to the differences in the parameter spaces spanned by the two sets of simulations. In future work, we will construct our own emulator for the matter power spectrum in order to facilitate a direct comparison with the Mira--Titan emulator.

\section{Data Release}
\label{sec:datarelease}
Upon posting of this article we are making the simulations described here available upon request. This includes the initial conditions, the particle snapshots and halo catalogs at all 10 redshifts described in \autoref{sec:simulations} and any measurements used in this paper or \citet{McClintock2018} and \citet{Zhai2017}. 

We will make the aforementioned data products freely downloadable at \url{https://AemulusProject.github.io} at the time this study and its companion papers are published.

\section{Discussion and Conclusions}
\label{sec:summary}
We have presented a new suite of $N$-body simulations for emulating cosmological observables. The cosmologies of these simulations were sampled from the $w$CDM $4\sigma$ allowed CMB+BAO+SN parameter space using an orthogonal Latin Hypercube. We investigated the convergence of the following observables with respect to choices made in the \textsc{L-Gadget2} \nbody\ solver: 
\begin{enumerate}[label=\alph*)]
\item Matter power spectrum, $P(k)$,
\item 3-dimensional matter correlation function, $\xi_{mm}(r)$,
\item Spherical overdensity halo mass function, $N(M_\text{200b})$
\item 3-dimensional halo--halo correlation function, $\xi_{hh}(r)$,
\item Projected galaxy--galaxy correlation function, $w_{p}(r_{p})$, and
\item Monopole and quadrupole moments of the redshift space galaxy--galaxy correlation function, $\xi_0(s)$, $\xi_{2}(s)$.
\end{enumerate}

We conclude that our observables are converged with respect to choices made in time stepping and force resolution. Choices with respect to initial conditions lead to minor deviations from $1\%$ convergence for halos resolved with fewer than 200 particles. Our choice of force softening leads to deviations from $1\%$ convergence for scales $r<200~\hkpc$. 

Particle loading is by far the parameter that our observables are most sensitive to. For halos with greater than 500 particles we also find convergence at better than the $1\%$ level, but for masses smaller than this, deviations from convergence due to insufficient particle loading increase rapidly. Halos with more than 200 particles, like those used in \citet{McClintock2018} are still converged to better than $2.5\%$. We have shown that this deviation is largely a function of particle number alone, and have fit this dependence and applied it to build the emulator in \citet{McClintock2018}. Additional tests in that study using even higher resolution simulations provide more evidence that this correction is satisfactory for our needs. 

We have shown that our halo mass function predictions are not affected by finite box effects to the precision allowed by sample variance in our test boxes. At $z=0$, sample variance is smaller than $1\%$ for $M_{200m}\sim < 4\times 10^{14} \hmsun$. The study of this effect in the current work is limited by our inability to use the same initial conditions for the different box sizes necessary for this test, as was done for the rest of the internal convergence tests detailed in this work. As such, these tests are limited by the sample variance in our test boxes, which is greater than percent level for $M_{\odot} \ge 4\times 10^{14} M_{\odot}$ and $z>$0. Future efforts are required to ensure that observables are indeed converged with respect to simulation size to the level needed for upcoming surveys. 

The matter power spectra in our simulations are consistent with those predicted by the \textsc{Halofit} methodology to within their reported errors, but we are unable to compare to the Mira--Titan emulator as our simulations use a different cosmological parameter space. Tests of this nature are of the utmost importance, and continued work in making them is vital to ensure that emulators of this kind are put to their full use in upcoming analyses.

The work presented here is just the beginning of our effort to contribute high precision and accuracy simulations and emulators to the community. Future work will extend our simulation suite significantly, especially with higher resolution simulations that are suited for use with more complete galaxy formation models. Additionally, we plan to expand our parameter space by including more physics such as neutrino masses, and by expanding the limits of the parameters sampled to include more volume away from CMB+BAO+SN constraints. This is important, as upcoming analyses will attempt to diagnose tension between different data sets in addition to combining constraints from many different experiments. 

The sharing of resources between simulators and the exchange of expertise between simulators, theorists, and observers will be vital in attaining the best possible outcomes for the next generation of surveys. Only a concerted effort from many groups in the domain of cosmic emulation over the next decade will help ensure that Stage-IV cosmological surveys are not limited by modeling systematics. 

\acknowledgments
This work received support from the U.S. Department of Energy under contract number DE-AC02-76SF00515. JLT and RHW acknowledge support of NSF grant AST-1211889. TM and ER are supported by DOE grant DE-SC0015975. ER acknowledges additional support by the Sloan Foundation, grant FG-2016-6443. YYM is supported by the Samuel P.\ Langley PITT PACC Postdoctoral Fellowship.
This research made use of computational resources at SLAC National Accelerator Laboratory, and the authors thank the SLAC computational team for support. This research used resources of the National Energy Research Scientific Computing Center, a DOE Office of Science User Facility supported by the Office of Science of the U.S. Department of Energy
under Contract No. DE-AC02-05CH11231.

\software{Python,
Matplotlib \citep{matplotlib},
NumPy \citep{numpy},
SciPy \citep{scipy},
nbodykit \citep{nbodykit},
Corrfunc \citep{corrfunc},
CCL \citep{ccl},
CAMB \citep{CAMB},
CLASS \citep{class},
}

\bibliographystyle{yahapj}
\bibliography{astroref,software}

\end{document}